\makeatletter
\newcounter{subhyp} 
\let\savedc@hyp\c@hyp

\newcommand{\normhyp}{%
  \let\c@hyp\savedc@hyp 
  \renewcommand\thehyp{\arabic{hyp}}%
} 

\documentclass[12pt]{article}
\usepackage{mwe}
\usepackage{subcaption}
\usepackage{amsmath, amsfonts, amssymb}
\usepackage{bm}
\usepackage{adjustbox}
\usepackage{amscd}
\usepackage{extarrows}

\usepackage{framed}
\renewenvironment{framed}[1][\hsize]
  {\MakeFramed{\hsize#1\advance\hsize-\width \FrameRestore}}%
  {\endMakeFramed}

\usepackage{moreverb,url}
\usepackage{xcolor}
\usepackage{graphicx}
\usepackage{caption}
\usepackage{subcaption}
\usepackage{amsmath}
\usepackage[colorlinks,bookmarksopen,bookmarksnumbered,citecolor=blue,urlcolor=blue]{hyperref}
\usepackage{setspace}   
\usepackage{authblk}
\usepackage[margin=1in]{geometry}
\usepackage{ntheorem}
\theoremseparator{:}

\usepackage[numbers,sort&compress]{natbib}
\usepackage{adjustbox}

\providecommand{\keywords}[1]
{
  \textbf{\textit{Keywords: }} #1
}

\title{{China's Rising Leadership in Global Science}}

\author[a,e,1]{Renli Wu}
\author[b,1]{Christopher Esposito}
\author[c,d,e,2]{James Evans}
\affil[a]{School of Information Management, Wuhan University, Hubei, China, 430072}
\affil[b]{Anderson School of Management, University of California, Los Angeles 90095}
\affil[c]{Department of Sociology, University of Chicago, IL 60615}
\affil[d]{Santa Fe Institute, NM 87501}
\affil[e]{Knowledge Lab, University of Chicago, IL, 60637}
\renewcommand\Affilfont{\itshape\small}

\begin{document}

\begin{titlepage}
\maketitle
\thispagestyle{empty} 
\begin{abstract}
\noindent
Major shifts in the global system of science and technology are destabilizing the global status order and demonstrating the capacity for emerging countries like China and India to exert greater influence. In order to measure changes in the global scientific system, we develop a framework to assess the hierarchical position of countries in the international scientific collaboration network. Using a machine-learning model to identify the leaders of 5,966,623 scientific teams that collaborated across international borders, we show that Chinese scientists substantially narrowed their leadership deficit with scientists from the US, UK, and EU between 1990 and 2023 in absolute terms. Consequently, China and the US are on track to reach an equal number of team leaders engaged in bilateral collaborations between 2027 and 2028. Nevertheless, Chinese progress has been considerably slower in per-collaborator terms: after adjusting for the number of non-leaders from each country, our models do not predict parity between the US and China until after 2087. These dynamics extend to 11 critical technology areas central to ongoing diplomacy between the two nations, such AI, Semiconductors, and Advanced Communications, and to China's scientific leadership with respect to the European Union and the United Kingdom. Thus, while China's elite scientists are achieving leadership in the international scientific community, China's scientific enterprise continues to face developmental constraints. We conclude by reviewing several steps that Chinese science is taking to overcome these constraints, by increasing its engagement in scientific training and research in signatory nations to the Belt and Road Initiative.\end{abstract}


\def\thefootnote{1}\footnotetext{These authors contributed equally to this work}
\def\thefootnote{2}\footnotetext{Corresponding author. Email: jevans@uchicago.edu}
\end{titlepage}

\doublespacing

\section{Introduction}

International research collaboration plays a pivotal role in scientific advance. Cooperation facilitates the sharing of specialized competencies, data, and resources, while promoting the global cross-pollination of ideas \cite{freeman2014collaboration}. International knowledge flows have become driving factors for economic growth and competitiveness \cite{hidalgo2007product,freeman2015immigration}, as science increasingly relies on large teams capable of performing heterogeneous tasks \cite{wuchty_increasing_2007, jones_burden_2009,haussler_2020}. Despite these benefits, international collaboration in science is a delicate affair. Members of scientific teams contribute different resources, creativity, and knowledge, a role differentiation that imparts hierarchical structure to the international collaboration network. Nations with scientific and technological superiority are more likely to position their scientists in leadership roles, steering the direction of scientific research. In contrast, scholars from countries still developing their scientific capacity tend to conduct supportive roles, granting them less command over science's direction and outputs. The imbalances that arise from the international division of scientific labor risk endangering global participation in international collaboration, and the enormous benefits it generates.

The United States has enjoyed nearly a century of leadership in global science, having obtained the position from Germany in the early 20th century \cite{xie_chinas_2014}. The influx of research immigrants from all over the world greatly benefited the U.S.\cite{moser2014german}, sustaining its status and expanding its output. Concurrently, China has undergone a process of rapid development and research ascension. As the world's second-largest economy, deeply integrated into the global trade system upon joining the WTO in 2001, China has also swiftly and successfully integrated into international networks of scientific collaboration over the past two decades, while continuously augmenting its investments in scientific research. China now outperforms the US in training more PhDs in the natural sciences and publishing more articles in top-ranked Nature-indexed journals \cite{baker_chian_2023}.

These developments have occurred during an era of increasingly complex international relations and a rise in anti-globalization sentiment and policy. While recent crises, including COVID-19, underscored the need for international research collaboration \cite{lewis2021research, beachy2023us}, the deteriorating political climate is eroding scientific ties and cooperation. Increasingly, scientists of Chinese descent in U.S. universities are experiencing increasing fear and anxiety, impacting their collaboration and career choices \cite{xie_chinas_2014}. Furthermore, a recent report from the U.S. Department of State shows a significant decline in the number of international students from China in U.S. higher education institutions, from 373,000 in 2019-2020 to 289,000 in 2022-2023 \cite{OpenDoors}. 

While China's economic growth has recently slowed, it continues to expand investment within its science and technology ecosystem. In addition, China is endeavoring to play an active role within international collaboration networks. Through increased investments in human capital, Chinese scientists are engaging more extensively with scientists in Central Asia, South Asia, and Sub-Saharan Africa, fostering the development of scientific enterprises in those regions. These investments could lay the groundwork for a "Silk Road of Science," extending China's sphere of scientific influence, broadening its cooperative partnerships worldwide, and potentially providing an additional opportunity for its many scientists to exert leadership in international scientific collaboration. 

In this paper, we report findings on China’s progress in achieving international scientific leadership, with respect to the US, UK, and the EU, and with respect to China's international partners that are signatories to the Belt and Road Initiative. We used a machine learning model to predict the leaders of 5,966,623 scientific teams that collaborated across the borders of "global regions" (such as China and the EU), and identified the share of team leaders involved in such collaborations that were based at Chinese institutions. By studying changes in China's share of international team leaders across time, we generated forecasts for when China is expected to reach parity in terms of their propensity to lead in collaborations with scientists from the US, EU, UK, and signatory nations to the Belt and Road Initiative. In addition to our main analysis, we also examine administrative data from the Chinese Ministry of Education to uncover investments made by China in enhancing its scientific collaboration and talent cultivation in Belt and Road Initiative signatory countries.

\section{China’s Leadership in International Teams}

\begin{figure}
\begin{framed}
    \centering
     \begin{subfigure}[!h]{0.48\textwidth}
         \centering
         \includegraphics[width=\textwidth]{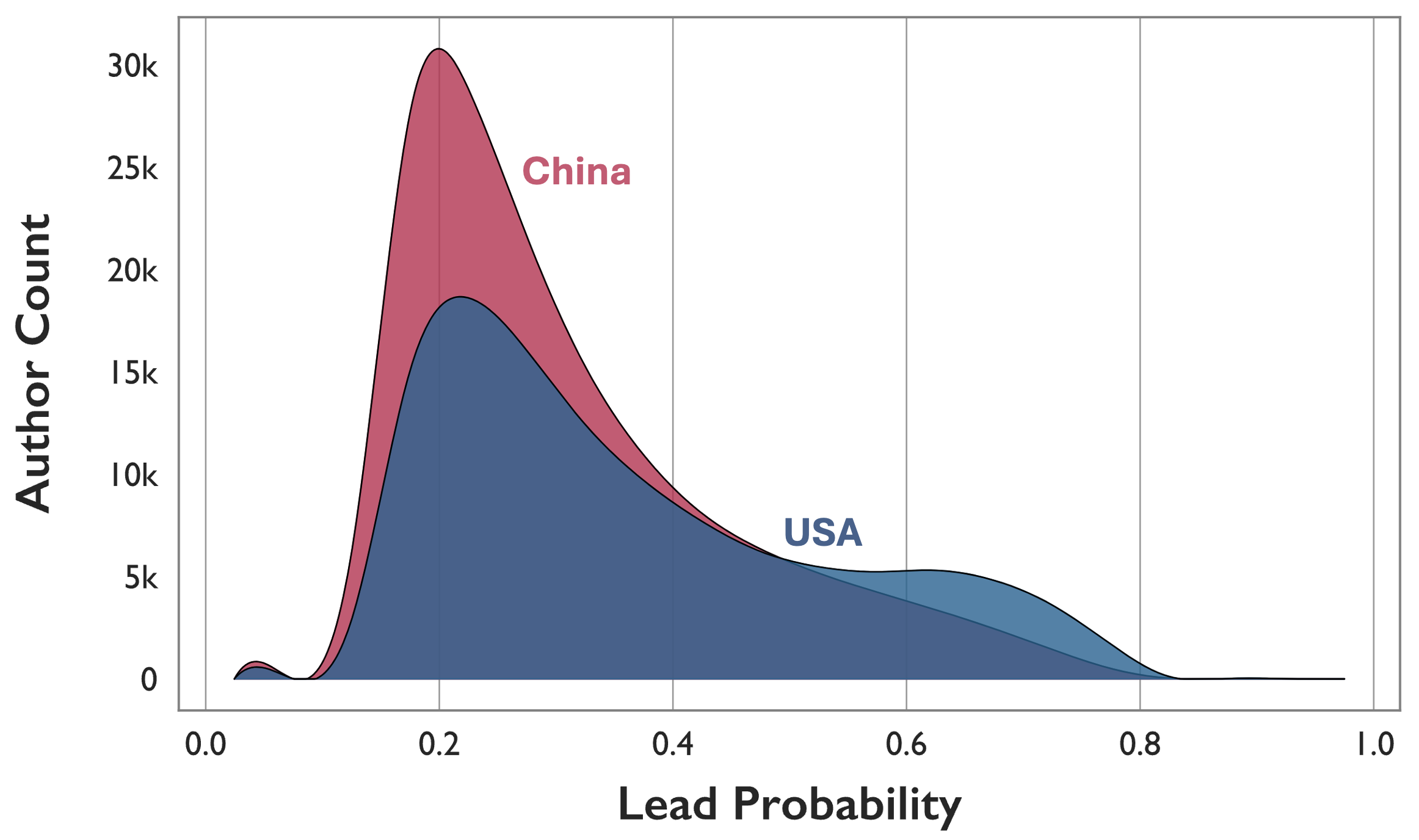}
         \caption{China-US Collaborators, 2010-2012}
         \label{fig:fig1a}
     \end{subfigure}
     \begin{subfigure}[!h]{0.48\textwidth}
         \centering
         \includegraphics[width=\textwidth]{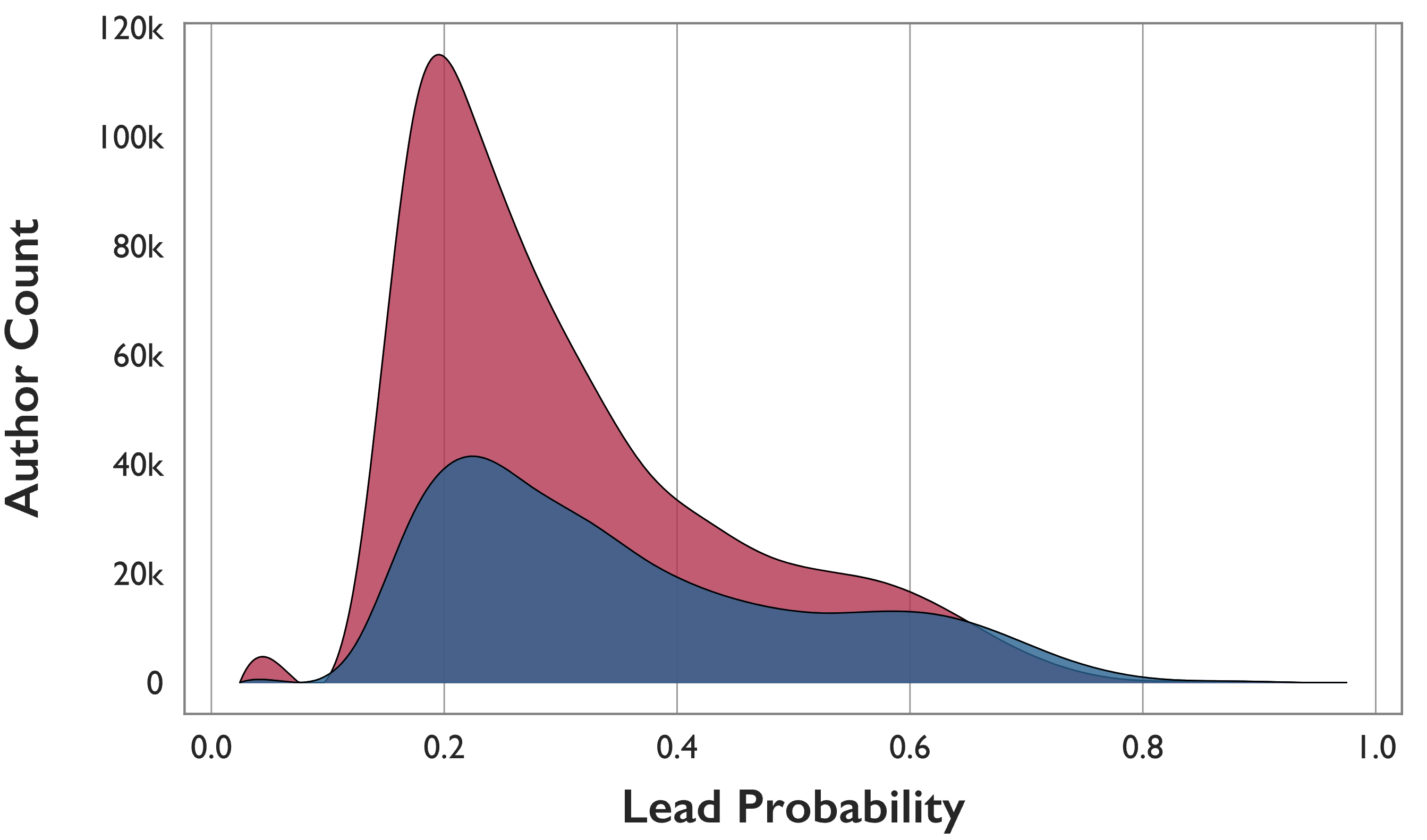}
         \caption{China-US Collaborators, 2021-2023}
         \label{fig:fig1b}
     \end{subfigure}
     \begin{subfigure}[!h]{0.48\textwidth}
         \centering
         \includegraphics[width=\textwidth]{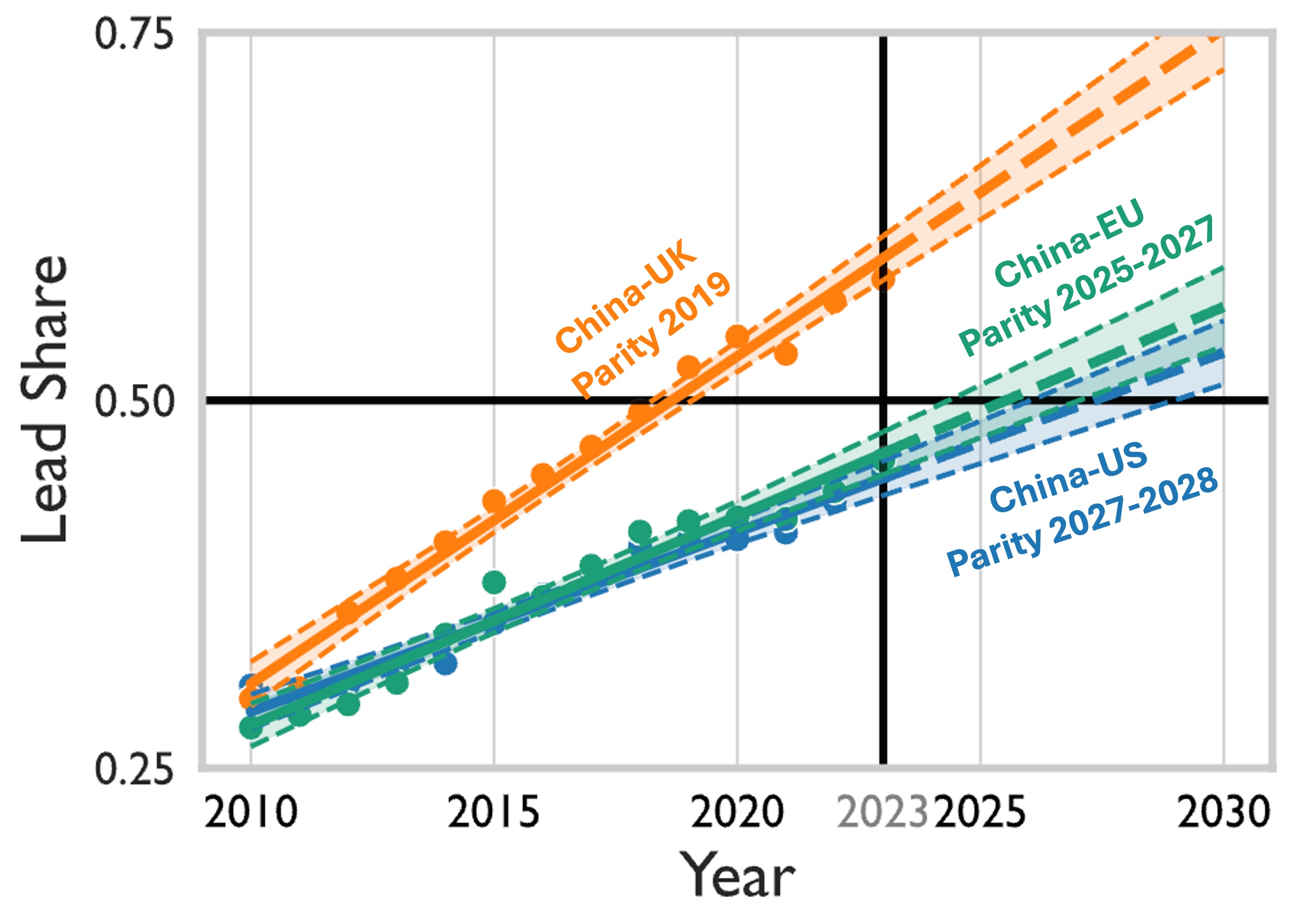}
         \caption{China's Lead Share with Select Partners}
         \label{fig:fig1c}
     \end{subfigure}
          \begin{subfigure}[!h]{0.48\textwidth}
         \centering
         \includegraphics[width=\textwidth]{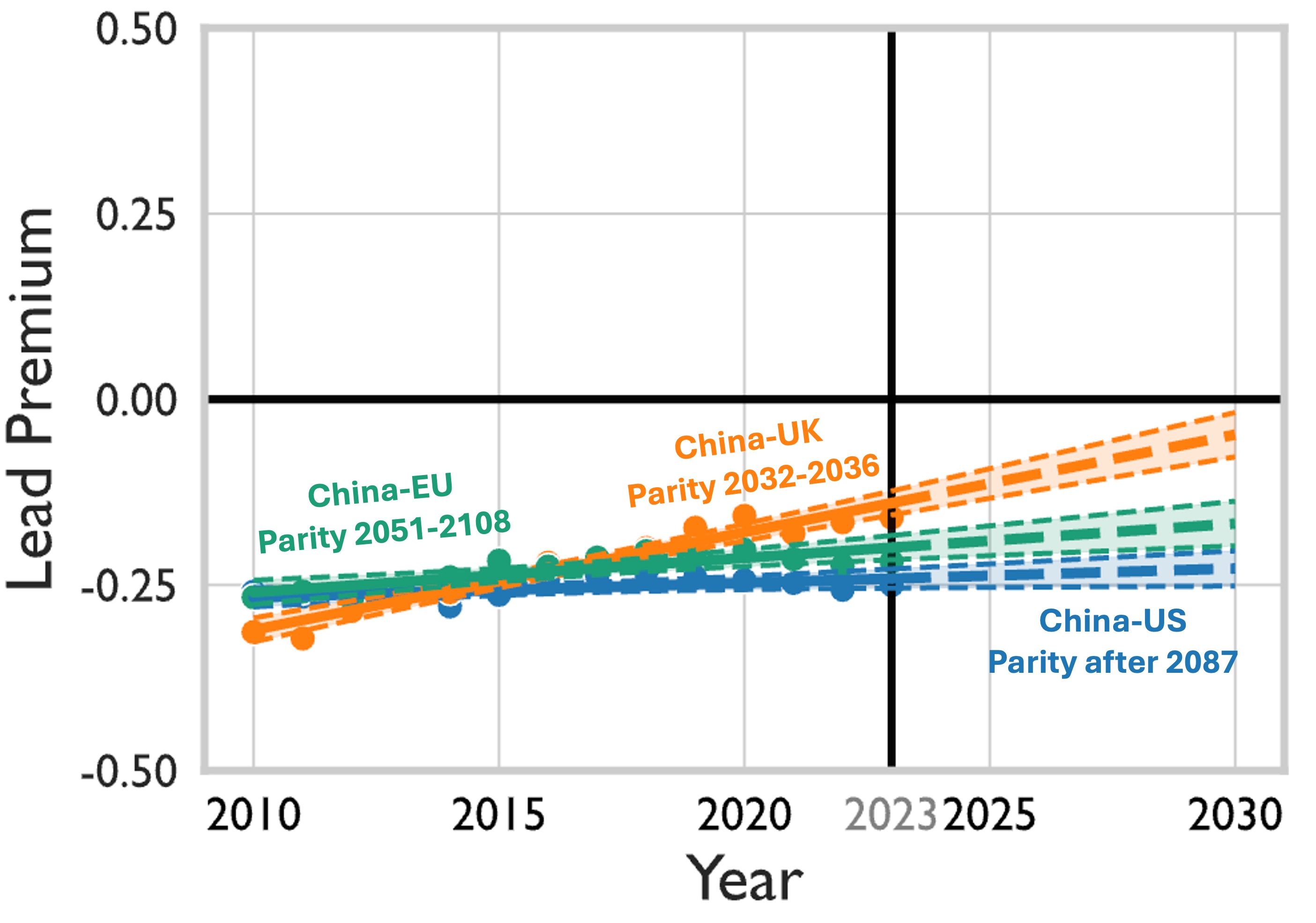}
         \caption{China's Lead Premium with Select Partners}
         \label{fig:fig1d}
     \end{subfigure}
\end{framed}
\caption{The Evolution of China's Leadership in International Scientific Collaborations}
\end{figure}
To assess China’s leadership status in the international collaboration network, we adopted the machine learning model developed by \cite{xu_flat_2022}, and data derived from publication contribution statements, to identify leaders within the nearly 6 million scientific publications indexed by OpenAlex that involved bilateral collaboration between 13 global regions, including China, the US, the UK, and the EU (Table S1). We find that Chinese scientific leadership vis-á-vis each of these three main partner regions increased substantially over our study period in absolute terms. Early in our study period (2010-2012), China had far fewer scientists with high lead probabilities involved in collaborations with the US than did the US with China (Figure \ref{fig:fig1a}). Nevertheless, this relationship changed by the end of the study period (2021-2023), when China had more scientists at moderate lead probabilities than did the US, and has substantially caught up to the U.S. at the highest lead probabilities (Figure \ref{fig:fig1b}). 

To quantify changes in Chinese scientific leadership across time, we define scientists with lead probabilities greater than 0.65 as leaders, and all scientists with lead probabilities below this threshold as supporters. We use these measures to compute China's \textit{Lead Share} with the partner regions, given by the share of leaders involved in China-US, China-EU, and China-UK collaborations that worked at Chinese-based institutions in each year (Figure \ref{fig:fig1c}). In 2010, about 30\% of the leaders in China-US collaborations were in China and 70\% were in the US. By 2023, 45\% were in China and 55\% were in the US. The relationship is approximated by a linear regression ($y$=0.012$x$-24.25), which is overlaid with 95\% confidence intervals in Figure \ref{fig:fig1c}. The upper bound of the confidence interval predicts parity between the US and China in 2027, while the lower bound predicts parity in 2028. For China and the EU, our model predicts parity between 2025-2027. For China and the UK, our model indicates that China reached parity in 2019.

China's rapid catch-up to the US, EU, and UK scientific leadership in absolute terms, as captured by the Lead Share, belies its potentially slower catch-up in per-collaborator terms. We computed China's \textit{Lead Premium} with each of the partner regions, which expresses a country's leadership in per-collaborator terms. The Lead Premium is calculated as a country's Lead Share minus its Supporter Share, and measures the extent to which China's collaborators with each partner country are biased toward leadership roles as opposed to supportive ones. We find that China's Lead Premium is not on track to reach parity with the US until after 2087, and that China is expected to reach parity with the EU between 2051 and 2108 and with the UK between 2032 and 2036 (Figure \ref{fig:fig1d}). Therefore, China continues to face challenges in shifting its large number of supporter scientists into leadership positions in international collaboration teams.

\section{Journal Quality and Leadership Threshold}

\begin{figure}
\begin{framed}
    \centering
     \begin{subfigure}[!h]{0.48\textwidth}
         \centering
         \includegraphics[width=\textwidth]{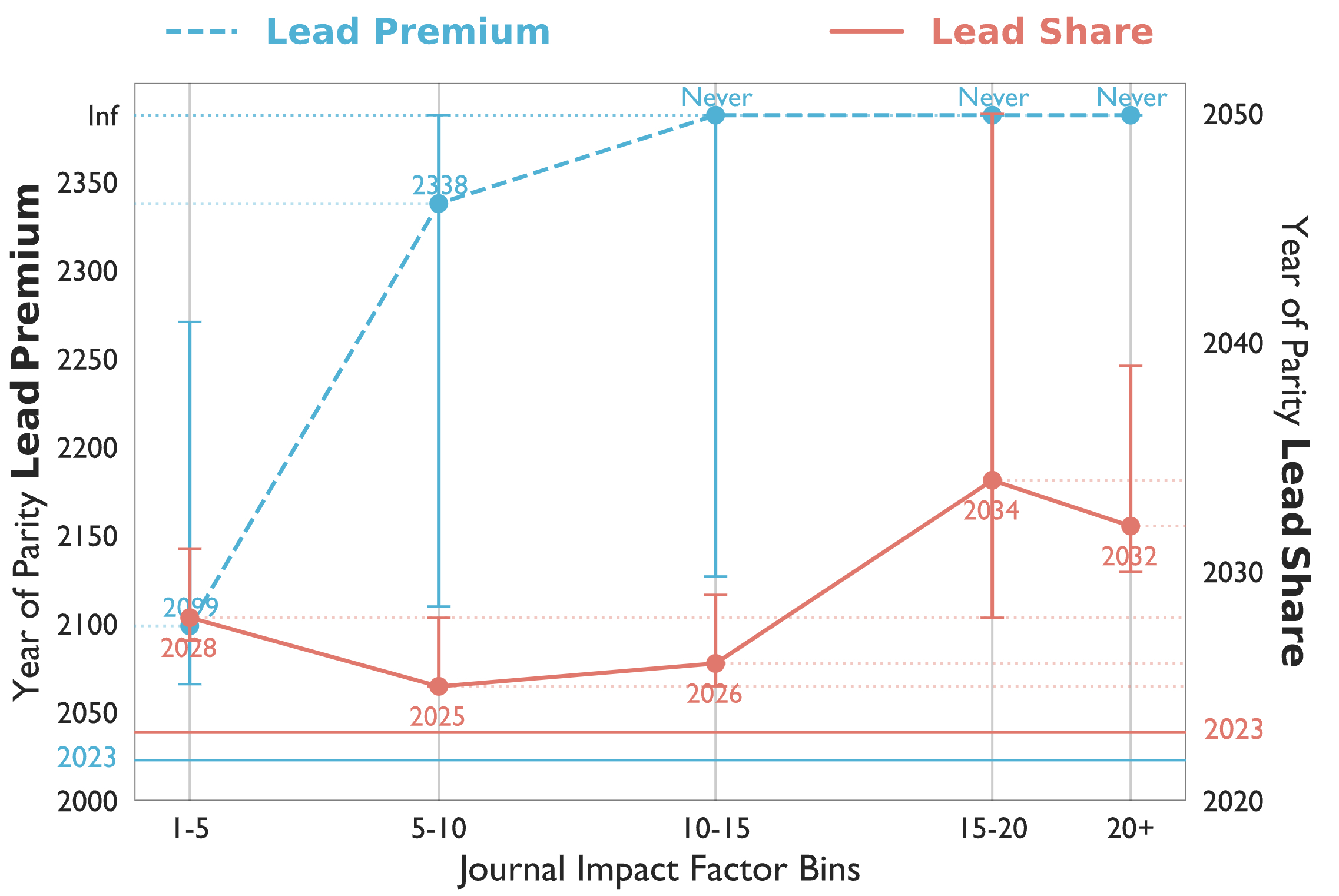}
         \caption{Parity Year by Journal Impact Factor}
         \label{fig:fig2a}
     \end{subfigure}
     \begin{subfigure}[!h]{0.48\textwidth}
         \centering
         \includegraphics[width=\textwidth]{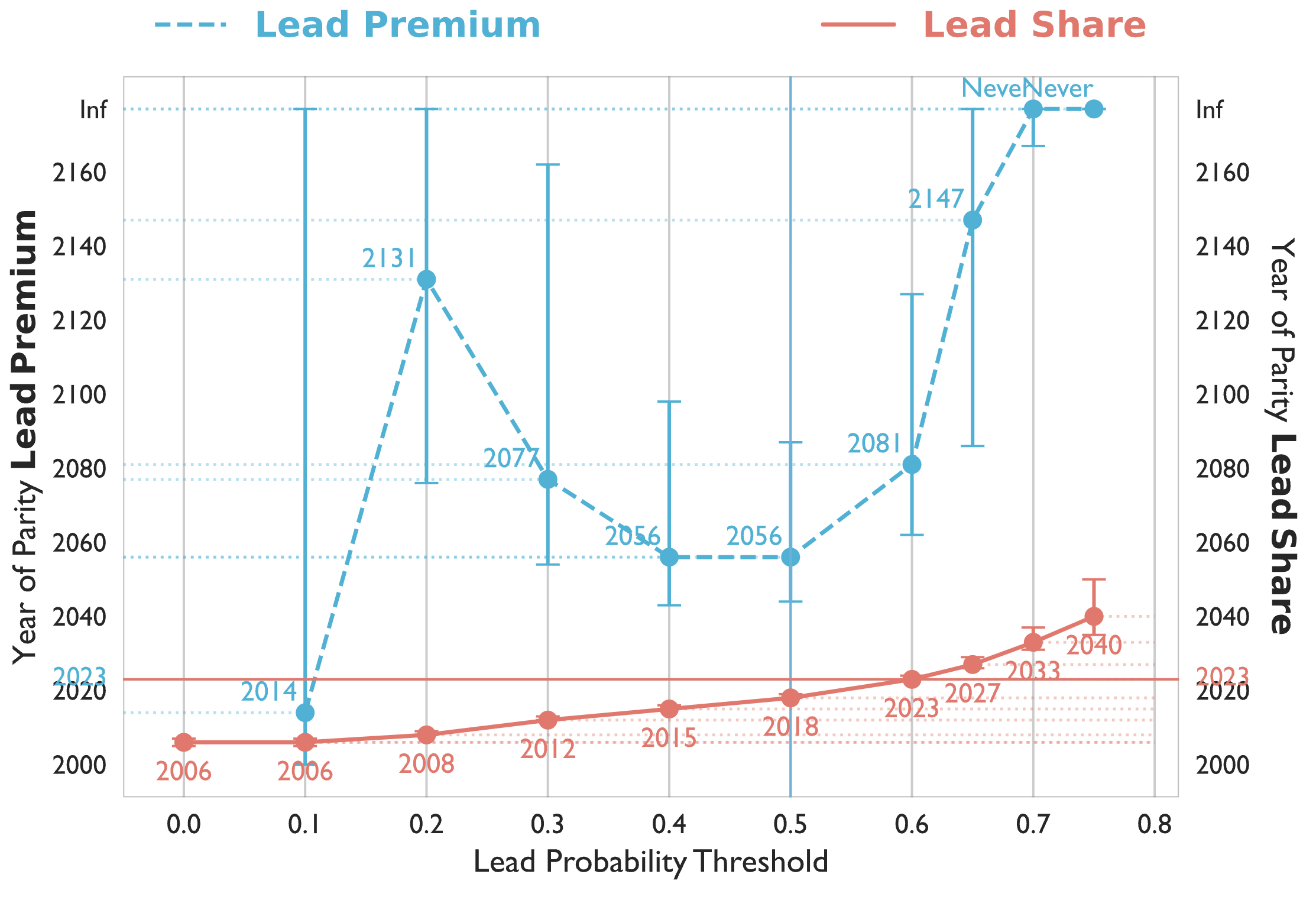}
         \caption{Parity Year by Lead Probability Threshold}
         \label{fig:fig2b}
     \end{subfigure}
\end{framed}
\caption{Year of Parity by Lead Journal Impact and Leadership Threshold}
\end{figure}

A view often expressed is that Chinese scientists may publish in lower-quality journals than do scientists from the US, EU, and UK. To test whether China's growing international leadership may be restricted to low-quality journals, we subsetted our data by journals based on their 2021 Web of Science Impact Factor. Figure \ref{fig:fig2a} shows predicted years of parity, with 95\% confidence intervals, for China's Lead Share and Lead Premium with the US across 5 impact factor journal bins. The point estimates for the Lead Share predict that China will reach parity with the US no later than 2034, irrespective of journal impact factor, while point estimates for Lead Premium predict that China will never reach parity with the US in high-impact journals. These results underscore that China's most elite scientists are becoming international leaders in the top journals, but that substantial challenges remain in transitioning many of its supportive scientists to leadership roles.

A second possibility regarding the emergence of Chinese scientific leadership is that China’s growth may be restricted to "mid-level leaders", while Western countries like the US might retain top-level leaders. To explore this possibility, we ran our model separately with different thresholds to define leaders. The point estimates predict that China will reach parity in Lead Share with the US at even the highest thresholds within the next two decades (Figure \ref{fig:fig2b}). In terms of its Lead Premium, Chinese parity is further off, with parity never expected between the two countries at the highest thresholds. 

\section{Chinese Leadership in Critical Technologies}

\begin{figure*}
\begin{framed}
\centering
\includegraphics[width=1\linewidth]{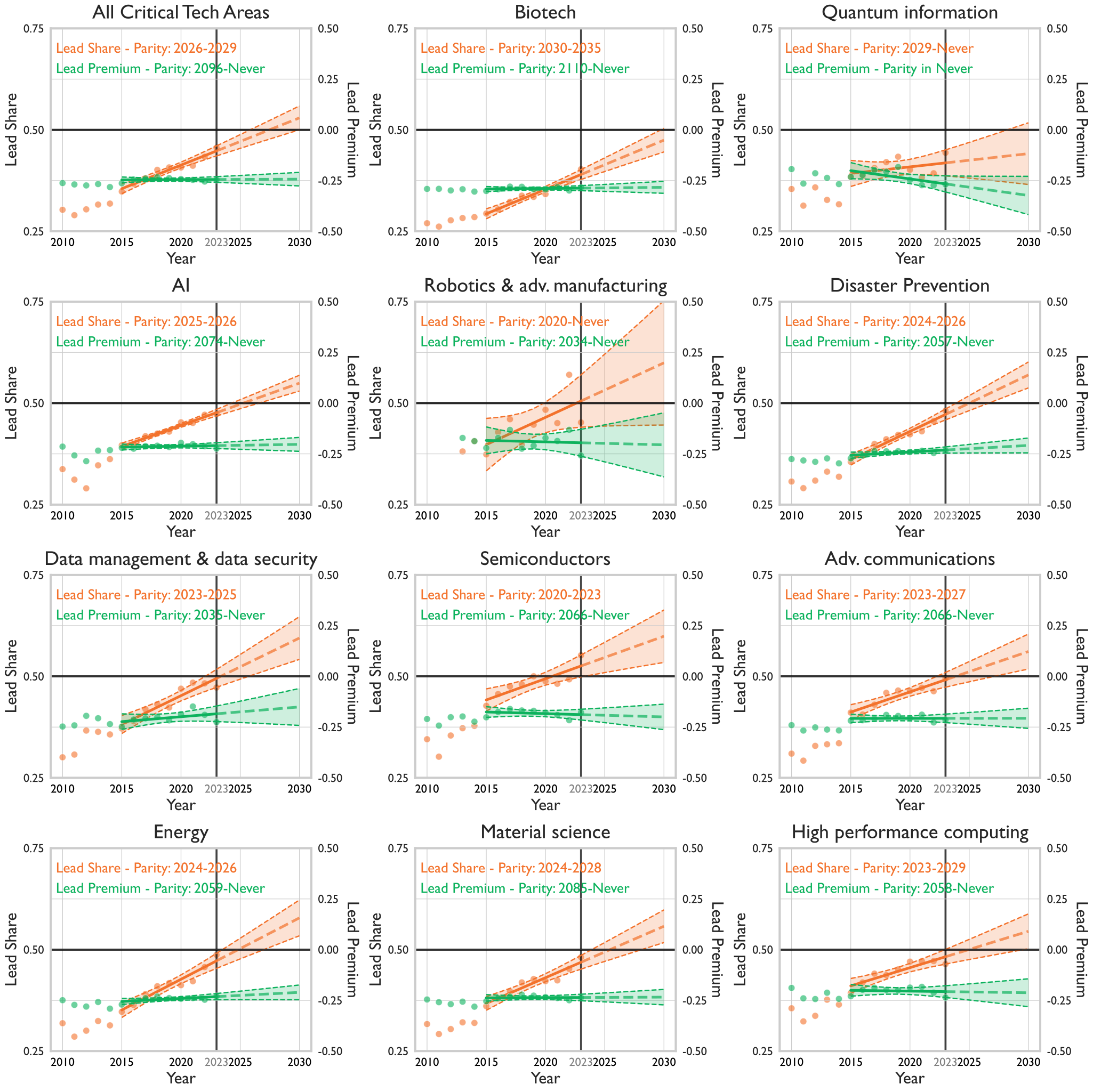}
\end{framed}
\caption{China’s Lead Share and Lead Premium with the U.S. in Critical Technological Areas}
\label{fig:fig3}
\end{figure*}

Global interest in scientific leadership is linked to the centrality of science for economic production, the provision of IT infrastructure, and maintenance of security. In the United States, the National Science Foundation's TIP Directorship defined 11 Critical Technology areas that it sees as essential to securing national objectives. These Critical Technology areas include Advanced Manufacturing, Advanced Materials, Artificial Intelligence, Biotechnology, Communications and Wireless, Cyberinfrastructure and Advanced Computing, Cybersecurity, Disaster Risk and Resilience, Energy Technology, Quantum Technology, and Semiconductors and Microelectronics. To assess the status of China-US leadership in these critical technologies, we ran our model separately for articles in each area (see methods).

Figure \ref{fig:fig3} shows China’s Lead Share and Lead Premium in bilateral scientific cooperation with the US over time in each of the 11 Critical Technology areas, plus a composite category. China is predicted to reach parity in its Lead Share with the US in the composite category between 2026-2029. Additionally, China is predicted to reach parity in its Lead Share with the US in 8 of the 11 Critical Technology Areas (AI, Disaster Prevention, Data Management and Security, Semiconductors, Advanced Communications, Energy, Materials Science, and High Performance Computing) before 2030. In Biotechnology, parity is not expected until 2030-2035, and in Quantum Information, parity is not expected until after 2029. In one field (Robotics and Advanced Manufacturing), our data is sparse, leading to noisy predictions. Finally, In terms of the Lead Premium, parity between the US and China is much further out, and not expected until after 2095 in the composite category.

In addition to the 11 Critical Technology Areas, we also analyzed our results by 6 broad scientific fields (Figure S12). We find that China is on track to reach parity in terms of its Lead Share with the US soonest in Chemistry and Materials Science (parity between 2023 and 2025), Mathematics, Physics, and Engineering (parity between 2023 and 2026), and Computer Science (parity between 2024 and 2026), and latest in Medicine (parity between 2031 and 2037). As in our other analyses, China's parity in its Lead Premium with the US is forested to be substantially further out.

\section{China’s Silk Road of Science}

Beyond the US, EU, and UK, we also found that China is engaging more deeply and broadly in scientific collaboration and talent development with the signatory countries of the Belt and Road Initiative. These developments would expand the scientific capacities of the signatory nations, while providing China with advantages associated with scientific leadership, including strengthening their mutual trust in science and technology and mobilizing more labor and resources toward their shared research objectives.

\begin{figure}
\begin{framed}
    \centering
     \begin{subfigure}[!h]{0.45\textwidth}
         \centering
         \includegraphics[width=\textwidth]{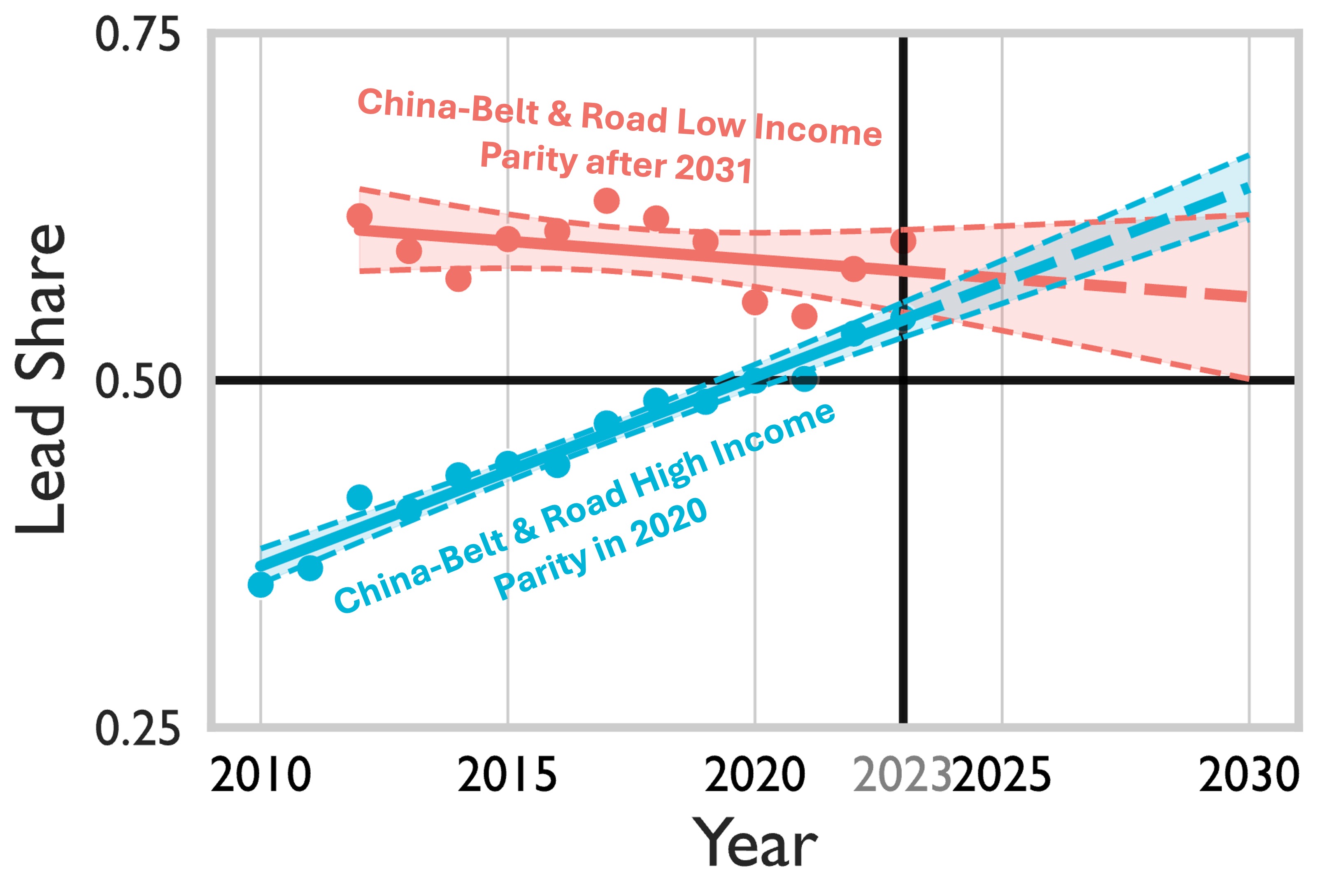}
         \caption{China's Lead Share with Belt and Road Countries}
         \label{fig:fig4a}
     \end{subfigure}
     \begin{subfigure}[!h]{0.45\textwidth}
         \centering
         \includegraphics[width=\textwidth]{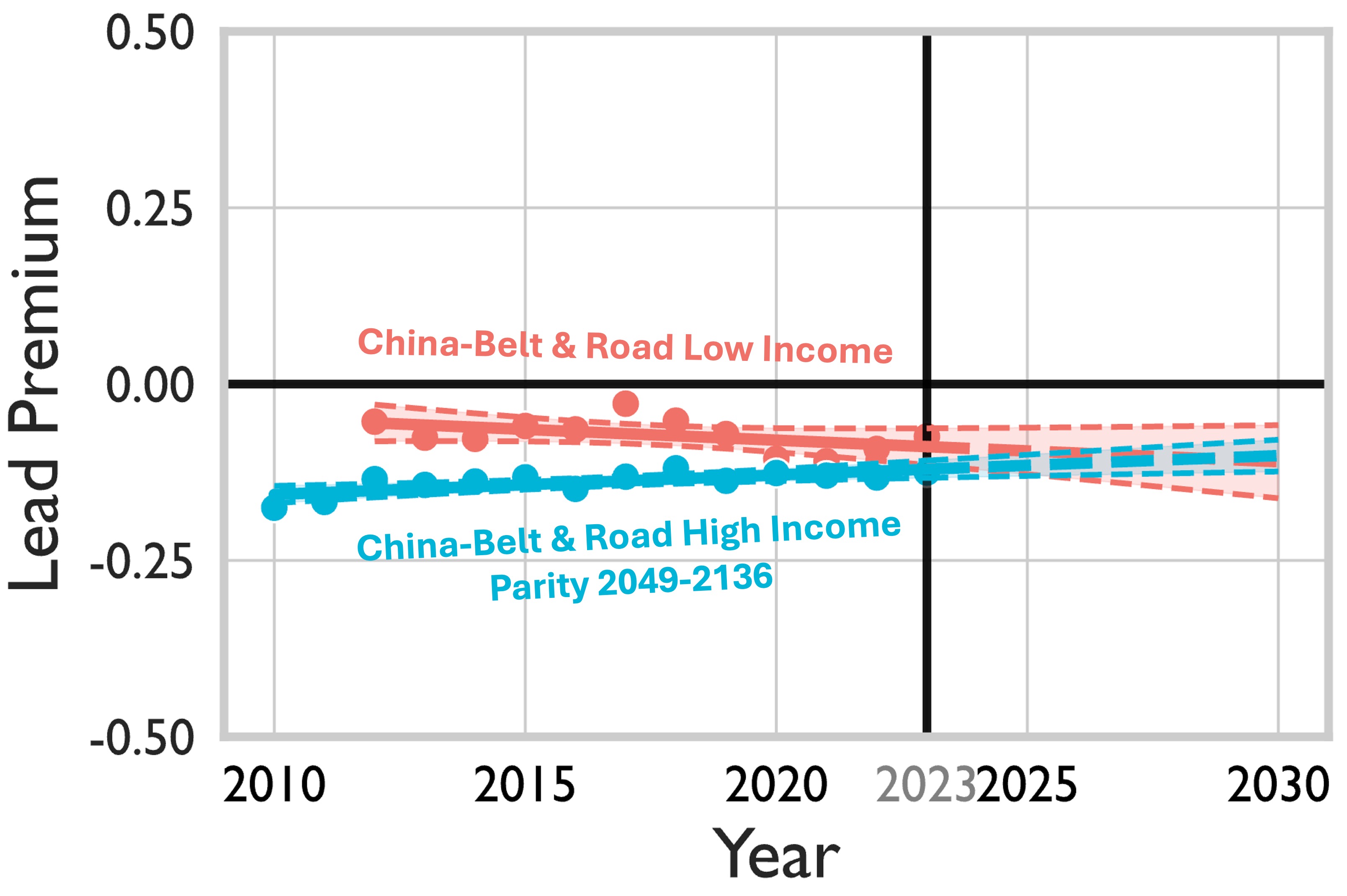}
         \caption{China's Lead Premium with Belt and Road Countries}
         \label{fig:fig4b}
     \end{subfigure}
     \begin{subfigure}[!h]{0.45\textwidth}
         \centering
         \includegraphics[width=\textwidth]{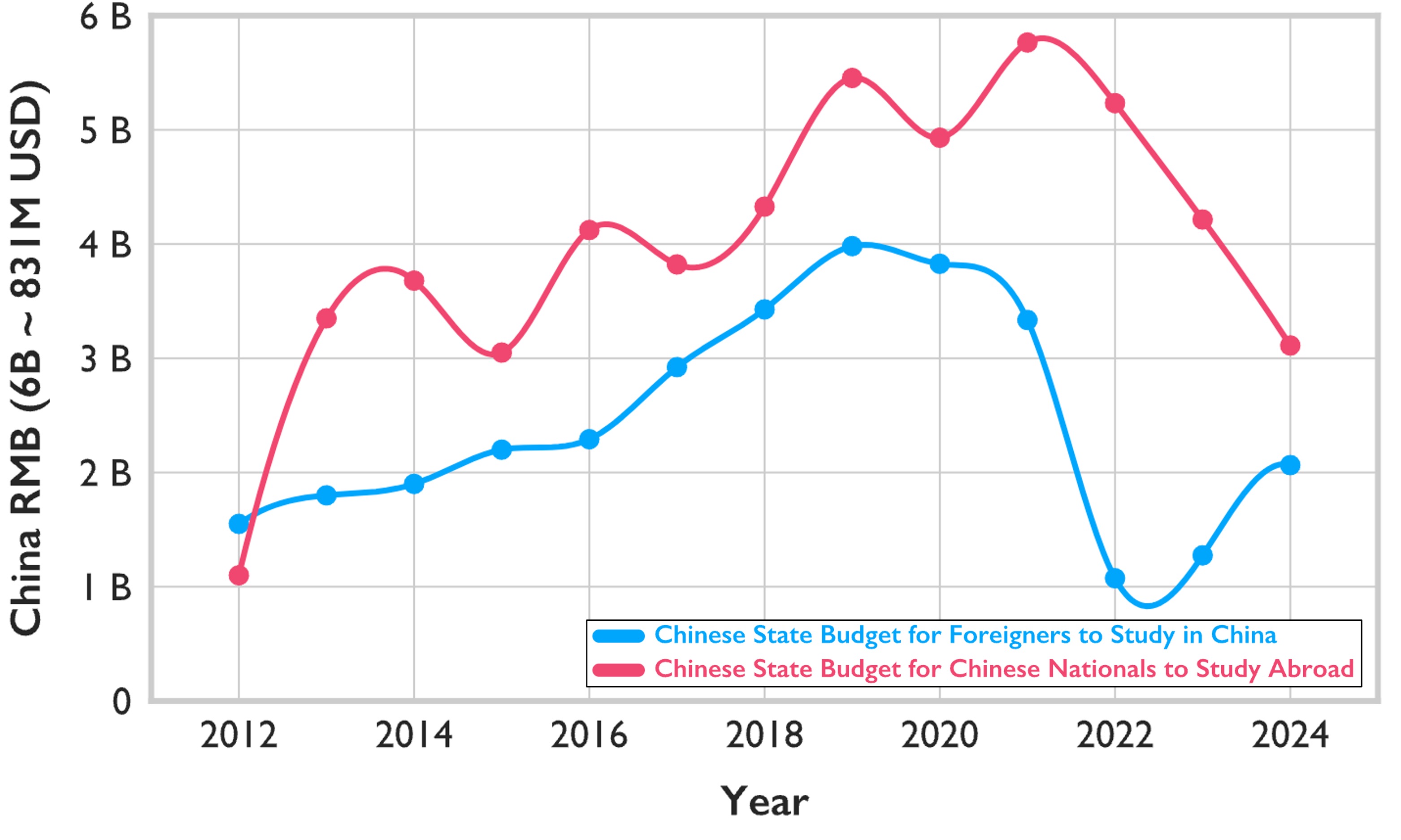}
         \caption{China's Government Budget for International Study}
         \label{fig:fig4c}
     \end{subfigure}
          \begin{subfigure}[!h]{0.45\textwidth}
         \centering
         \includegraphics[width=\textwidth]{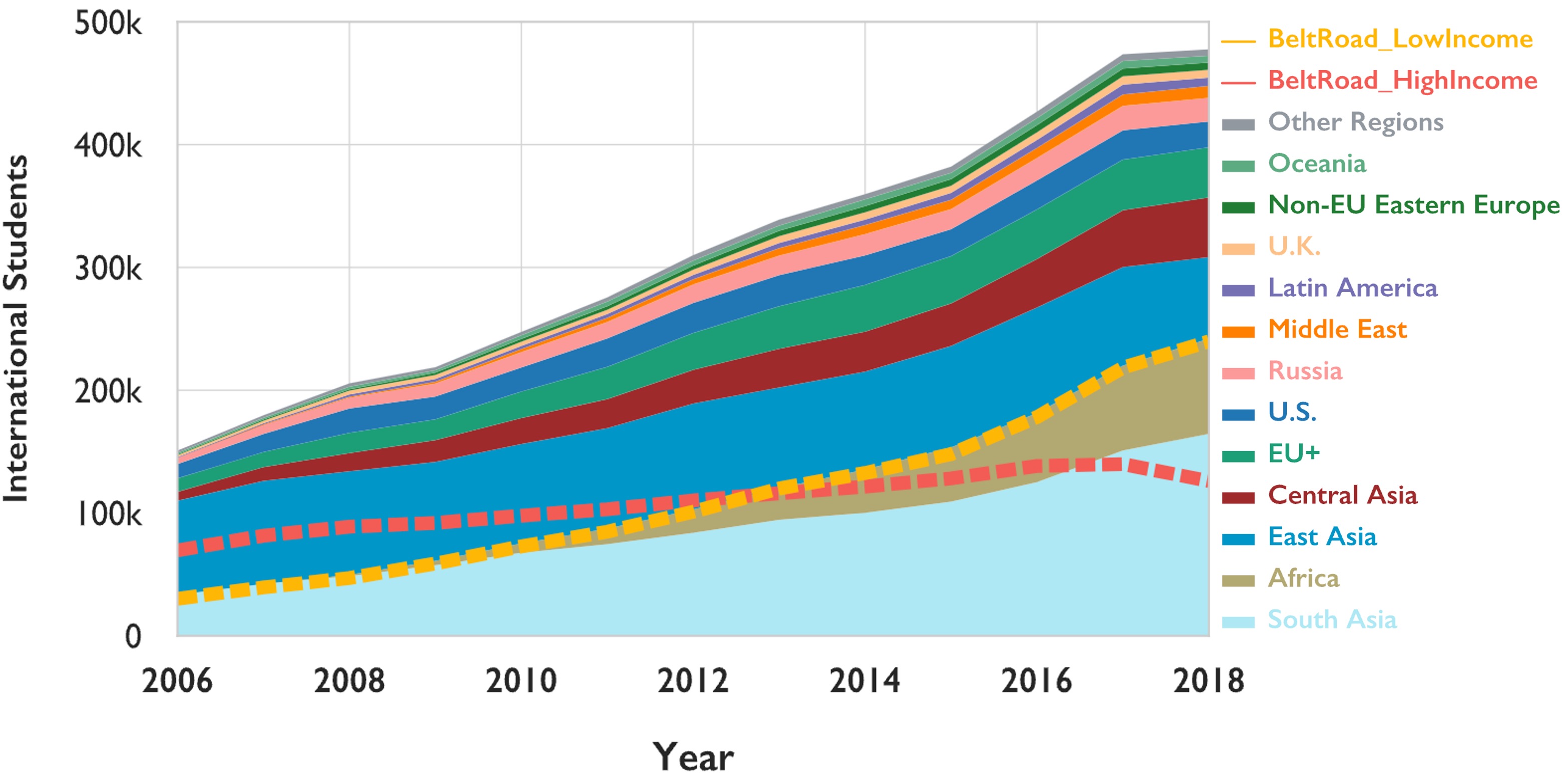}
         \caption{Foreign Students in China by Origin}
         \label{fig:fig4d}
     \end{subfigure}
          \begin{subfigure}[!h]{0.7\textwidth}
         \centering
         \includegraphics[width=\textwidth]{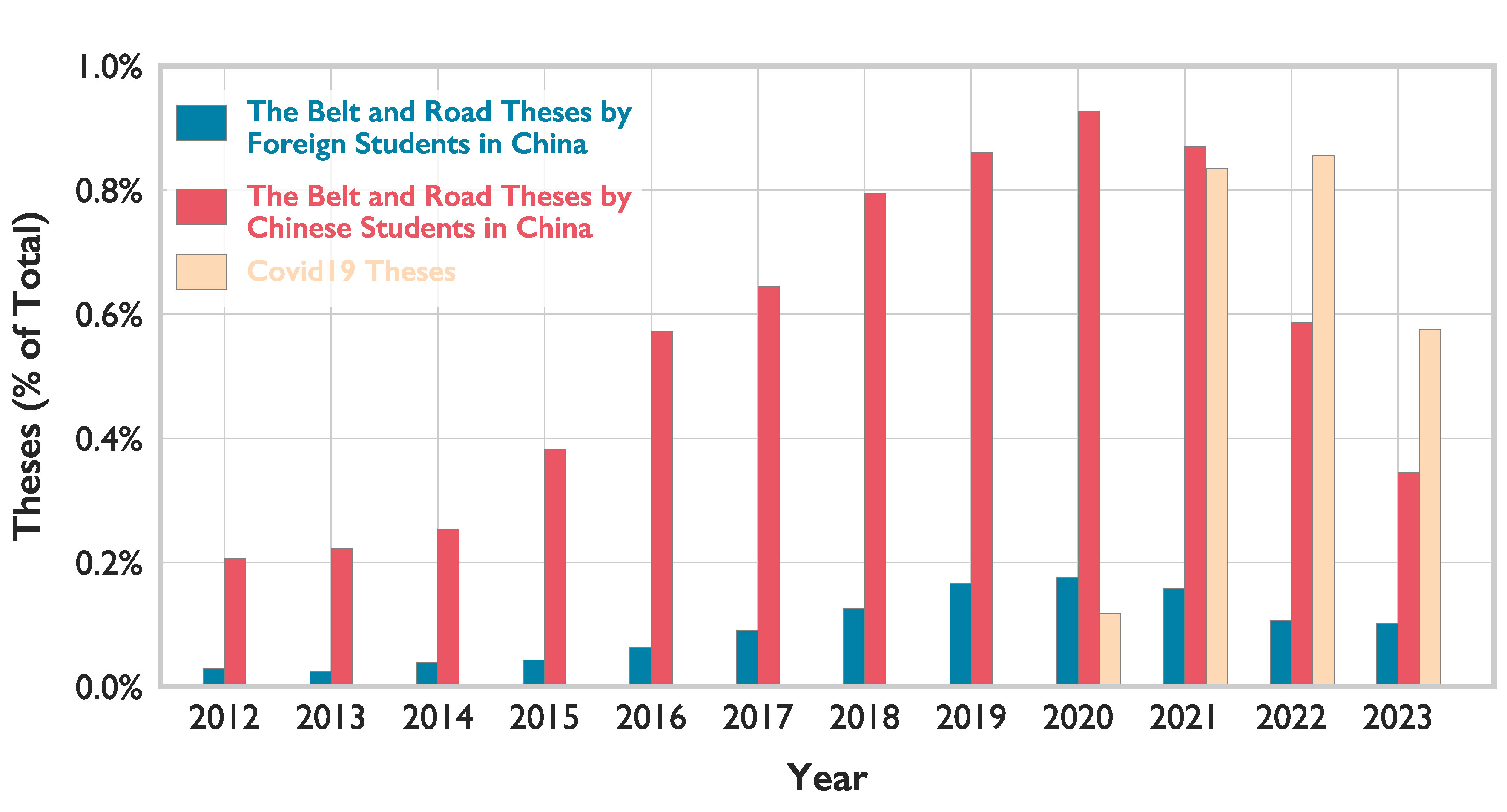}
         \caption{Thesis Topics in China}
         \label{fig:fig4e}
     \end{subfigure}
\end{framed}
     \caption{Year of Parity by Lead Journal Impact and Leadership Threshold}
\end{figure}

To study the growth of Chinese scientific leadership in Belt and Road countries, we first reviewed China's Lead Share and Lead Premium with the Belt and Road signatory countries. A diverse set of countries have joined the initiative, including very wealthy countries such as Italy, Austria, and New Zealand, as well as developing countries. Therefore, we divided Belt and Road Initiative countries into two groups: those with GDP per capital below that of China using 2010 nominal values, and those with GDP per capital above China's 2010 value. Figure \ref{fig:fig4a} shows that China's Lead Share is higher than that of the low-income Belt and Road countries, and that it crossed the parity line with high-income Belt and Road countries in 2020. Figure \ref{fig:fig4b} shows that parity in China's Lead Premium with both high and low-income Belt and Road countries looms in the near future. Therefore, the current trends in Chinese leadership with Belt and Road Initiative countries do not break from China's leadership status with respect to the US, EU, and UK: while China has elite scientists leading international collaborations, it has many scientists not on track to assume leadership roles at the international scale, without unseen interventions.

One potential pathway through which China could shift more of its scientists into international leadership positions is by training young scientists from Belt and Road countries at Chinese universities. We examined budgetary data from China’s Ministry of Education, which reveal that the Chinese government budgeted over 32 billion RMB (\$4.5 billion USD) between 2012 and 2024 to support foreign students’ study in China (Figure \ref{fig:fig4b}). This amount is only 30\% less than the amount China spent on its own nationals studying abroad, and is 66\% of the size of the total budget requested by the U.S. National Science Foundation in 2018. Most importantly, while spending for foreigners' study in China plummeted during the COVID-19 pandemic, it re-surged in 2023 and 2024, which suggests that China plans to continue these investments.

Student-level data from the Chinese Ministry of Education reveal the geographical origin of foreigners studying in China (Figure \ref{fig:fig4c}). In 2006, most of China's foreign students were from East Asia, with a smaller share from South Asia. The share from South Asia and Africa grew over time, and collectively accounted for nearly half of all of China's foreign students by 2018, the most recent year for which the Ministry has provided data. Dotted lines in Figure \ref{fig:fig4c} show the number of students from Belt and Road signatory countries that studied in China in each year, broken out by country income level. During the time series, China shifted from hosting more Belt and Road students from high-income countries, to hosting more Belt and Road students from low-income countries. Thus, China shifted its student inflows toward countries with which its own leadership may be more readily developed.

Finally, to identify the type of research produced by foreigners studying at Chinese universities, we conducted a content search through the full corpus of masters and doctoral theses in the China National Knowledge Infrastructure (CNKI) database. The CNKI database is one of the most comprehensive repositories of scholarly content published in China. We recorded the number of theses written by Chinese nationals and by foreigners studying in China that were related to the Belt and Road, as an indicator of the extent of research activity in China that is helping China to deepen its global integration. In 2012, 0.21\% of the theses written in China were about the Belt and Road, but by 2019 that figure rose to 1.17\%. More importantly, the number of theses about the Belt and Road written by foreigners studying in China rose six-fold during the same time period (Figure \ref{fig:fig4d}). Thus, through a combination of its education subsidies and rising scientific prominence, China is attracting immigrant students to its country, where they are engaging in research that promotes cultural exchange and deepens cultural and economic integration between their home nations and China.

\section{Discussion}

Our findings document the rapid growth of Chinese scientific leadership in the global collaboration network, as well as challenges China faces in shifting its population of scientists into international leadership. Analyzing the full corpus of bilateral collaborations between China, the US, and the EU, we find that China is on track to achieve a leadership position with Western countries in absolute terms within the next decade. These findings extend to leadership in high-impact journals, the most elite levels of leadership, and 8 of the 11 technological areas central to ongoing diplomatic discussions. Despite this progress, China's leadership growth is markedly slower in per-collaborator terms.

When considering policy action, these developments demand careful reflection. China and the United States benefit enormously from one-another’s excellence in science: scientists from the two countries collaborate more frequently than any other country or global-region pair, with the exception of the US and EU (Figure S1), and collaborations between China and the US disproportionately benefit science as a whole \cite{alshebli_china_2023}. Policies to isolate scientific research in either country would entail large costs to both countries, as well as the rest of the world that benefits from the outputs of their research. Additionally, both countries have already developed strong positions of scientific leadership with respect to one-another and other global regions. This limits the leverage either country has to impede progress of the other. 

For policymakers in the United States, China’s investment in scientific enterprises across Belt and Road Initiative countries inspires a different course of action: the US can deepen its investments in global human capital development. By expanding its scientific engagement in the developing world, through education, training, and scientific collaboration, the US can train a global workforce in the skills demanded by the US economy, facilitate cultural exchange, and strengthen its position as central pole of the global professional class.

Chinese policymakers may take note of China's increasing Lead Share with partner countries, which signifies that China's policies to develop its scientific capabilities have helped make the country's elite scientists globally competitive. Nevertheless, attention should also be directed at China's slower growth in its Lead Premium, which suggests that additional action could be taken to bolster leadership and innovation capabilities across its population of scientists. In this regard, China's recent investments in cultivating the scientific ecosystems in Belt and Road countries may prove a fruitful strategy, and could be expanded. Chinese policymakers may consider increasing investments to enhance the quality of talent cultivation. By leveraging the educational resources and scientific strengths of China's institutions as well as those of more developed Belt and Road countries, they can improve the educational levels of international students and expand training programs for professionals critically needed by less developed Belt and Road countries.

Most importantly, our findings suggest that the dynamics of U.S.-Chinese competition in scientific leadership can be shifted in a more productive and mutually-beneficial direction. Whether or not competition results in negative or positive outcomes depends on the nature of the competition. Efforts to restrain the scientific progress of either country is hazardous, and such efforts are unlikely to succeed given the sophistication of both countries' scientific enterprises. On the other hand, competition that takes the form of integrating each country's science more deeply into global scientific networks through investments in  human capital development would generate positive externalities for the world, by expanding global educational opportunities and accelerating scientific progress.

\section{Methods and Materials}

\subsection*{Publication Data}

We source publication data from the December 2023 data dump of OpenAlex, the largest repository of scientific publications in social, natural, physical, computational, and biological science journals. We sourced all published articles (5,966,627 publications) that fit the following criteria: (a) the publication date was after 1990, (b) the publication appeared in a journal with a 2021 impact factor of 1 or greater, and (c) the authors of the publication were affiliated with research/educational educations (universities, research institutes, or think tanks) located in two Global Regions, as defined below. This third criterion implies that we subsetted only the publications that involved bilateral collaborations between scientists in two separate Regions of Geopolitical Significance. For example, we do not analyze publications involving three scientists, one from the U.S., one from the EU, and one from China, because the collaboration is not bilateral. We plot the number of publications by each region pair in Figure S1.

\subsection*{Definitions of Global Regions}
We aggregated countries to Global Regions, based on their strategic orientation, funding regimes, and levels of scientific development. We defined 13 Global Regions: China (excluding Hong Kong Special Administrative Region and Taiwan Province), the United States, the United Kingdom, the European Union (the EU27 plus Switzerland and Norway), Russia, Non-EU Eastern Europe, Oceania, East Asia, South Asia, Central Asia, the Middle East, Africa, and Latin America. A full list of countries by their assigned geopolitical region is provided in Table S1. 

\subsection*{Identifying Team Leaders}
To identify leaders in teams of scientists, we computed the lead probability for each author in each paper. Lead probabilities range from 0 to 1, with higher values indicating a higher likelihood that the given scientist was a cognitive and creative leader the the creation of the publication. Lead probabilities are computed separately for each author for each of her or his publications, using the method of \cite{xu_flat_2022}. Specifically, we utilized the dataset provided by \cite{xu_flat_2022}, which parsed the 83,877 publications with self-reported author contribution statements from \textit{Science}, \textit{Nature}, \textit{PNAS}, and \textit{PLOS ONE}. These contributions were then clustered into lead-related contributions, indirect support-related contributions, and direct support-related contributions, as a function of the likelihood that each of these roles are held by the same person on the team (see \cite{xu_flat_2022}). Clustered leadership roles include the contribution statement activities: “conceive,” “design,” “lead,” “supervise,” “coordinate,” “interpret,” and “write.” Direct support roles include: “help,” “assist,” “prepare,” “develop,” “collect,” “generate,” “purify,” “carry,” “do,” “perform,” “conduct,” and “analyze.” Indirect support activities include “participate,” “provide,” “contribute,” “comment,” “discuss,” and “edit.” Using fractional assignment, authors were then assigned a lead value of 1 if they made lead-related contributions, and value of 0 if they made indirect support-related contributions or direct support-related contributions. Consequently, the resulting training dataset contains a list of all author-paper observations for the selected publications, each associated with their respective lead values according to the clustered contribution statements. We used 90\% of the dataset from 4 as the training dataset and withheld 10\% of the data for post-training model evaluation.

We trained the model using nine features: (1) number of references in the focal paper's bibliography previously cited by the focal author, (2) number of topic keywords appearing in the focal paper that previously appeared in an author’s prior publications, and (3) number of self-citations in the focal paper to the author’s prior work, (4) career age of the author; (5) number of prior publications by the author; (6) total number of citations received by the author leading up to the focal paper; (7) number of unique keywords explored by the author leading up to the focal paper; (8) number of past publications where the author ranked as the first author or the last author; and (9) the affiliation score based on the importance rank of the author. Using the withheld 10\% of the dataset, our model has a precision of 0.692 and a recall of 0.705.   

Our model differs from the one used by \cite{xu_flat_2022} in that ours does not consider author order or corresponding status when predicting lead probabilities (which is unreliable for international publications), ours includes self-citations and affiliation rank. See Figure S4 for the distribution of lead probabilities for the 5,966,623-paper dataset. We exploit the bimodal distribution of the lead probability distribution to create a binary classification of leaders. In the main analyses, we defined all author$\times$paper observations with lead probabilities $>$ 0.65 as “leaders” and all observations with lead probabilities $\leq$ 0.65 as “supporters”. In Figure \ref{fig:fig2b}, we showed results using multiple lead probability thresholds to create a binary leadership definition.

We computed the Lead Share on a pairwise basis, as the share of leaders that engaged in a bilateral collaborative team between two Global Regions that were from the focal Global Region. We computed the Lead Premium as each country's Lead Share minus its Supporter Share. Finally, we note that these two measures are complementary, and reveal different attributes about a Region's position in the global collaborative network. For example, China has a very large population of scientists, which naturally lends it to perform well in terms of its Lead Share (because only a small share of China's scientists need to lead internationally for the country to have a high Lead Share with partner Regions), but also to perform poorly in terms of its Lead Premium (because it would be very difficult for the majority of China's many scientists to establish leadership positions internationally). We thus report both measures, allowing for contextualized interpretation.





\bibliography{bib_mar_22_2024}

\end{document}



\maketitle

\SItext

\section*{Papers by Region Pair and Signatory Countries to the Belt and Road Initiative}

\begin{figure}
\centering
\includegraphics[width=\textwidth]{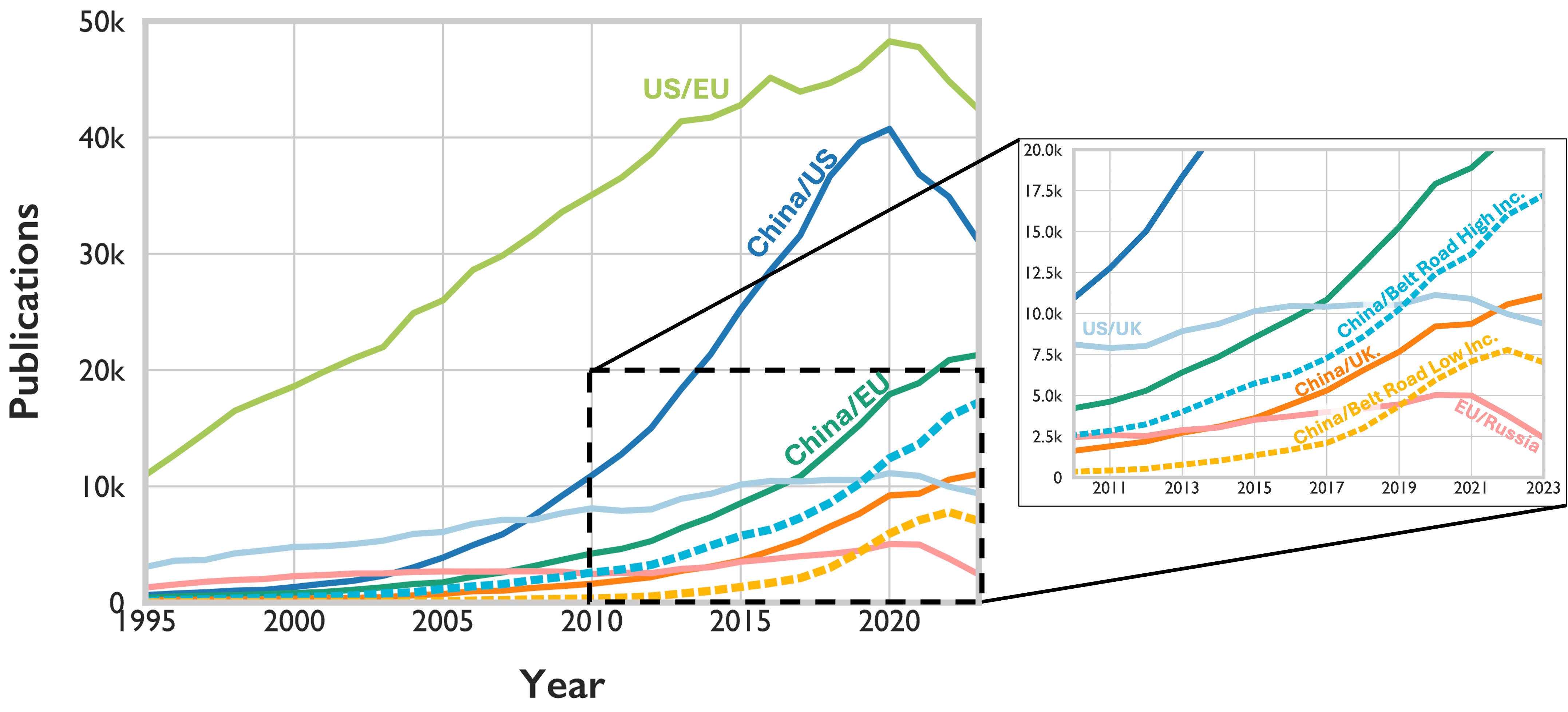}
\caption{Count of Publications by Region Pair}
\label{S1}
\end{figure}

Figure \ref{S1} shows the number of papers produced in bilateral collaborations between global regions for the full 5,966,627-paper OpenAlex dataset. The largest share of papers were produced through US-EU collaborations, but the fastest growth in papers were produced by US-China collaborations (until COVID-19 disrupted international collaboration). The insert image on the right shows that the number of collaborations between scientists from China and Belt and Road countries experienced rapid growth starting in 2010 (for high-income Belt and Road countries) and starting in 2015 (for low-income Belt and Road countries), and that by 2023, there were about 17,500 collaborations between China and high-income Belt and Road countries and about 7,500 collaborations between China and low-income Belt and Road countries. For reference, in 2023 there were just under 10,000 bilateral collaborations between the US and UK.

Table \ref{T1} shows how we assigned countries to Global Regions. Finally, Figure \ref{S2} contains a list of Belt and Road signatory countries, broken out by income level. We classified signatory countries by income based on their 2010 nominal GDP per capita. High-income countries are those with higher GDP per capita than China in 2010; low-income countries are those with lower GDP per capita than China.

\section*{Identifying Team Leaders}

We identified team leaders in the full dataset (5.9 million publications) using the methods described in the main text’s Methods section. Panel A of Figure \ref{S3}  illustrates the results of our clustering method, with different contribution types assigned to 3 unique clusters (two of which are both supporting roles). Panel B of Figure \ref{S3} shows the association between the 9 features we observe for each scientist and their lead probability.

Figure \ref{S4} is a density plot of the distribution of Lead Probability of all scientists in our 5.9 million publication dataset, broken out by team size. 

\section*{Changes in Lead Shares from US, EU, and UK perspectives}

In Figure 2 of the main text, we showed the Lead Share and Lead Premium of China for scientists engaged in collaborations with the US, UK, and EU. In addition, we fitted a linear regression model to the data from for the years 2010-2021 in order to predict the year of parity. In Figure 2, we also labeled the year at which China was expected to reach parity with the other three regions in terms of its Lead Share and Lead Premium.

In the supplement, we show similar results for the U.S. (Figure \ref{S5}), the EU (Figure \ref{S6}), and the UK (Figure \ref{S7}). The most notable new insight generated by these figures is taht the UK-centric figure shows that the UK is declining in its Lead Share and Lead Premium with all three other partners regions (the US, the EU, and China). 

\section*{Critical Technology Areas}

We studied US-Chinese parity year predictions (using our preferred outcome measure, the Lead Share) for the 11 Critical Technology areas defined by the NSF in the US as technologies that are core to their country’s economic production and defense systems. See Methods section for full details. Figure \ref{S8} provides a list of NSF-provided keywords for each of the Critical Technology areas. Figure \ref{S9} provides a list of the OpenAlex concepts that we assigned to each critical technology area. Figure \ref{S10} shows the number of papers in our full (5.9 million) paper dataset that involved bilateral collaborations between China and the US by Critical Technology Area.

\section*{Scientific Fields}

In addition to the 11 critical technology areas, we also analyzed our results across 6 broad scientific fields. The fields are: chemistry and materials science, computer science, earth and life sciences, humanities and social sciences, mathematics + physics + engineering, and medicine. We assigned papers to these fields based on their level 0 concepts from OpenAlex, which were generated using information from publication titles, abstracts, published journals and citation networks. Note that every paper appears in at least one scientific field. Figure \ref{S11} shows that collaboration between Chinese and U.S. scientists is most frequent in chemistry and materials science and second-most frequent in the earth and life sciences. Collaboration between Chinese and U.S. scientists is least frequent in medicine and in the humanities and social sciences.

Figure \ref{S12} shows China’s lead share with the US across the 6 technology fields. Notably, China expected to reach parity with the US across all fields within the next 15 years. The predicted parity range for Medicine is somewhat further out than in the other fields.

\section*{China’s Publications with Scientists from Belt and Road Countries}

In Figure \ref{S13}, we show the frequency of collaboration of Chinese scientists with scientists from Belt and Road by country income level and scientific field. 

\section*{Replication of Analysis using OpenAlex Data}

We replicated our analysis using data from Microsoft Academic Graph (MAG) using the 2021 Dec. data compilation for papers with impact factors greater than or equal to 1 (using 2021 impact factor scores) and found similar results. In Figure \ref{S14}, we show the number of papers in the MAG dataset by region pair for 1995-2021.

We also replicated our main analysis of trends in China’s leadership using the MAG dataset, as shown for China-US, China-UK, and China-EU in Figure \ref{S15}, and for China-Belt and Road Countries in Figure \ref{S16}. The MAG dataset uses a different name disambiguation for the scientists, so it is slightly different (in addition to having shorter and more limited journal coverage). Given these differences, it is notable that our main findings extend to the MAG dataset.


\begin{figure}
\centering
\textbf{High-Income Belt and Road Initiative Countries:} Antigua and Barbuda, Argentina, Austria, Bahrain, Bangladesh, Brunei Darussalam, Bulgaria, Chile, Cook Islands, Cuba, Cyprus, Czech Republic, Dominica, Dominican Republic, Equatorial Guinea, Estonia, Gabon, Greece, Grenada, Guyana, Hungary, Italy, Kazakhstan, Kuwait, Latvia, Lebanon, Libya, Lithuania, Luxembourg, Malaysia, Maldives, Malta, Montenegro, New Zealand, Oman, Panama, Poland, Portugal, Qatar, Romania, Russian Federation, Saudi Arabia, Seychelles, Singapore, Slovak Republic, Slovakia, Slovenia, South Africa, South Korea, Suriname, Trinidad and Tobago, Turkey, United Arab Emirates, Uruguay \textbf{Low-Income Belt and Road Initiative Countries:} Afghanistan, Albania, Algeria, Angola, Armenia, Azerbaijan, Barbados, Belarus, Benin, Bolivia, Botswana, Burundi, Cabo Verde, Cambodia, Cameroon, Chad, Comoros, Congo, Rep., Costa Rica, Cote d'Ivoire, Croatia, Democratic Republic of the Congo, Djibouti, Ecuador, Egypt, Arab Rep., El Salvador, Eritrea, Ethiopia, Fiji, Gambia, The, Georgia, Ghana, Guinea, Guinea-Bissau, Indonesia, Iran, Islamic Rep., Iraq, Jamaica, Kenya, Kiribati, Kyrgyz Republic, Lao PDR, Lesotho, Liberia, Madagascar, Malawi, Mali, Mauritania, Micronesia, Fed. Sts., Moldova, Mongolia, Morocco, Mozambique, Myanmar, Namibia, Nepal, Nicaragua, Niger, Nigeria, Niue, North Macedonia, Pakistan, Papua New Guinea, Peru, Philippines, Rwanda, Samoa, Senegal, Serbia, Sierra Leone, Solomon Islands, Somalia, South Sudan, Sri Lanka, Sudan, Syrian Arab Republic, Tajikistan, Tanzania, Thailand, Timor-Leste, Togo, Tonga, Tunisia, Turkmenistan, Uganda, Ukraine, Uzbekistan, Vanuatu, Venezuela, RB, Vietnam, Yemen, Rep., Zambia, Zimbabwe
\caption{List of Belt and Road Initiative Countries}
\label{S2}
\end{figure}

\begin{figure}
\centering
\includegraphics[width=\textwidth]{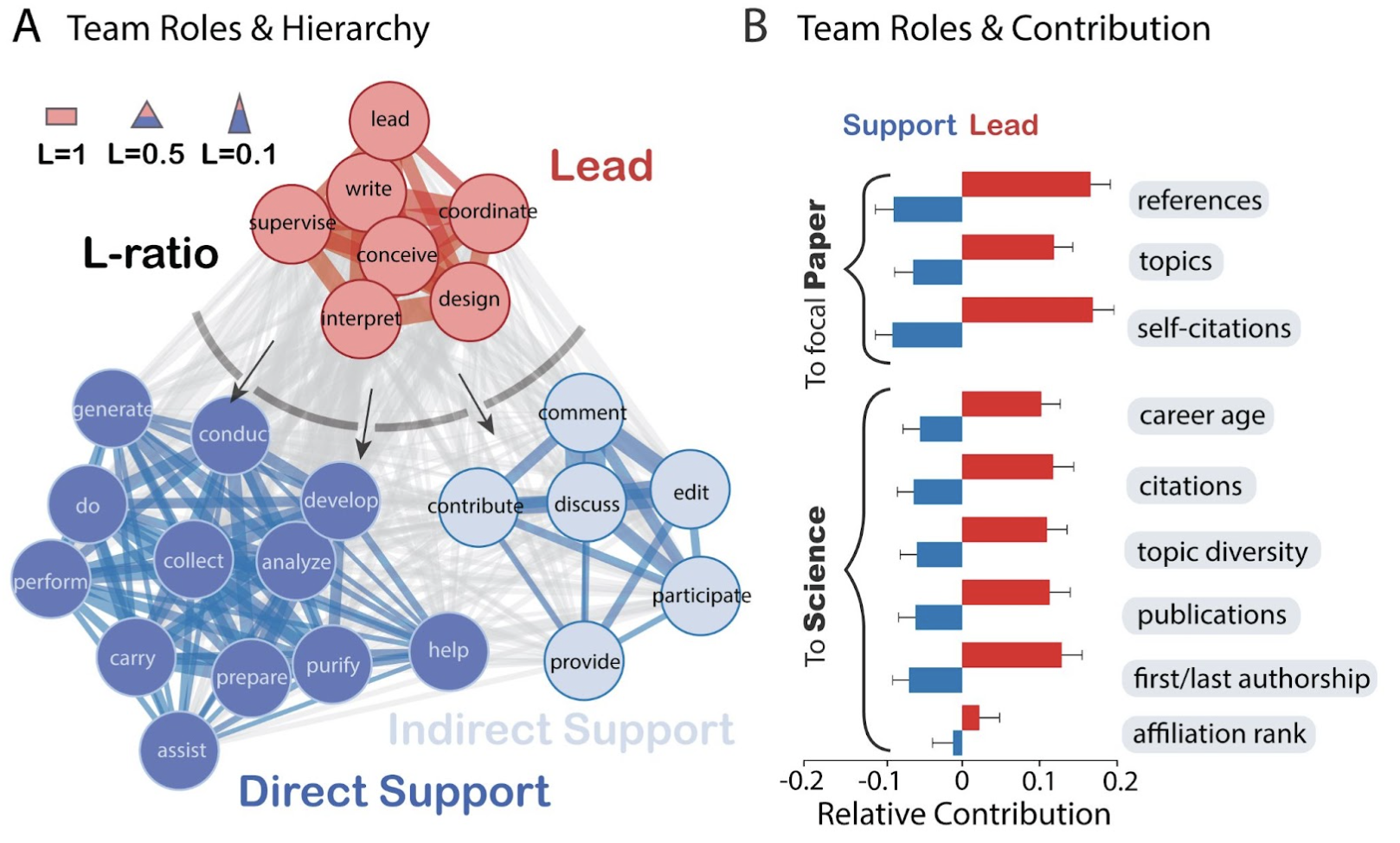}
\caption{Deep Learning Model for Identifying Lead Authors\\\textit{Note:} Figure S3A (left panel) shows the network of author contributions from the training dataset, clustered on basis of their co-ocurence within the same author and the same paper. Our k-means clustering identifies three unique clusters, which we manually label as Lead, Direct Support, and Indirect Support clusters based on the terms contained within each cluster. We then collapse Direct Support and Indirect Support functions into a single "Support" function to create a binary Lead/Support definition of task contributions. Figure S3B (right panel) shows coefficients from a linear model applied to the training data set that predicts the extent to which an author contributed leadership (as opposed to supportive) tasks, on basis of 9 author*paper features (see Methods section in the main text).}
\label{S3}
\end{figure}

\begin{figure}
\centering
\includegraphics[width=\textwidth]{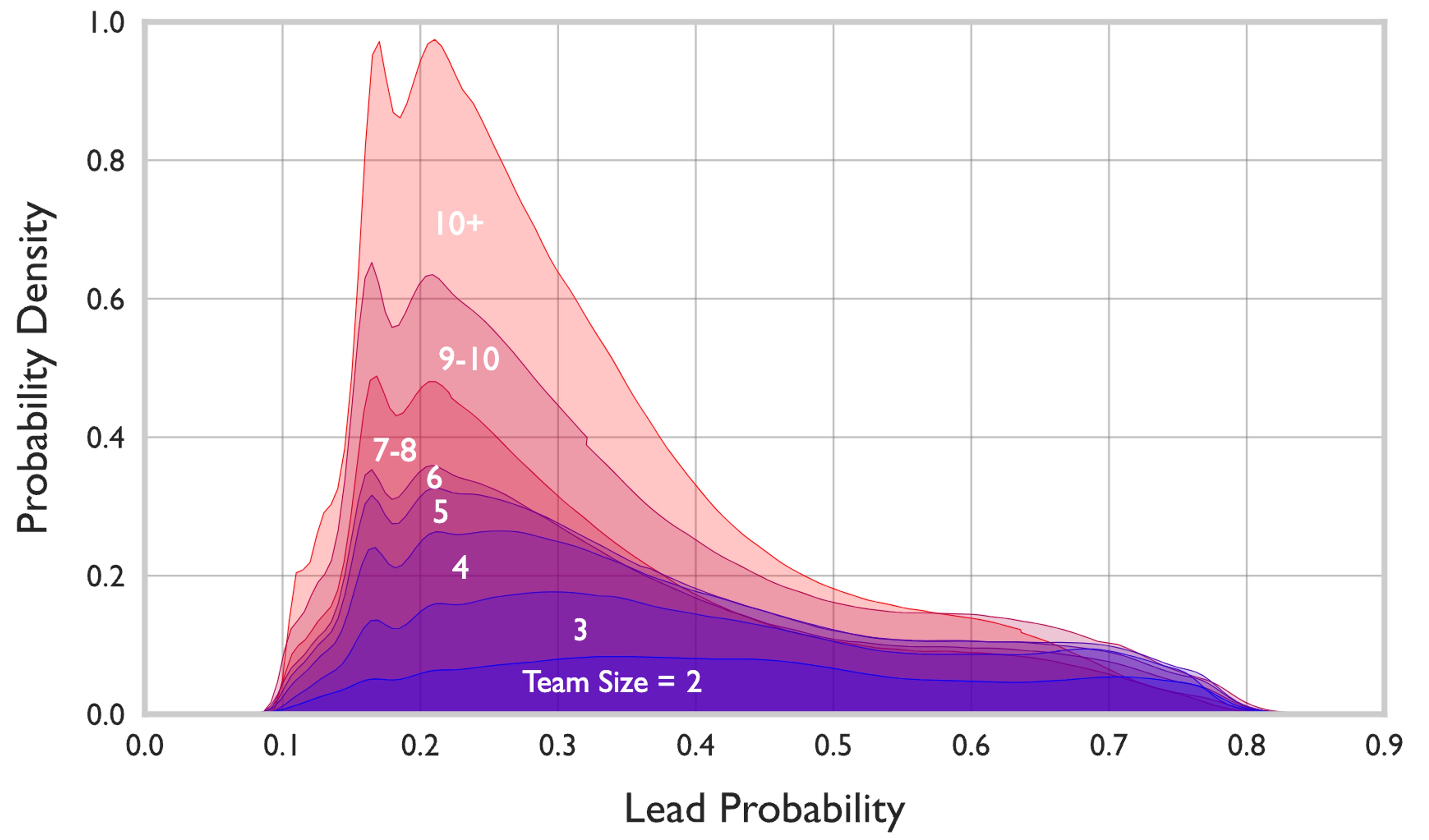}
\caption{Lead Probability Distribution by Team Size}
\label{S4}
\end{figure}

\begin{figure}
\centering
\includegraphics[width=\textwidth]{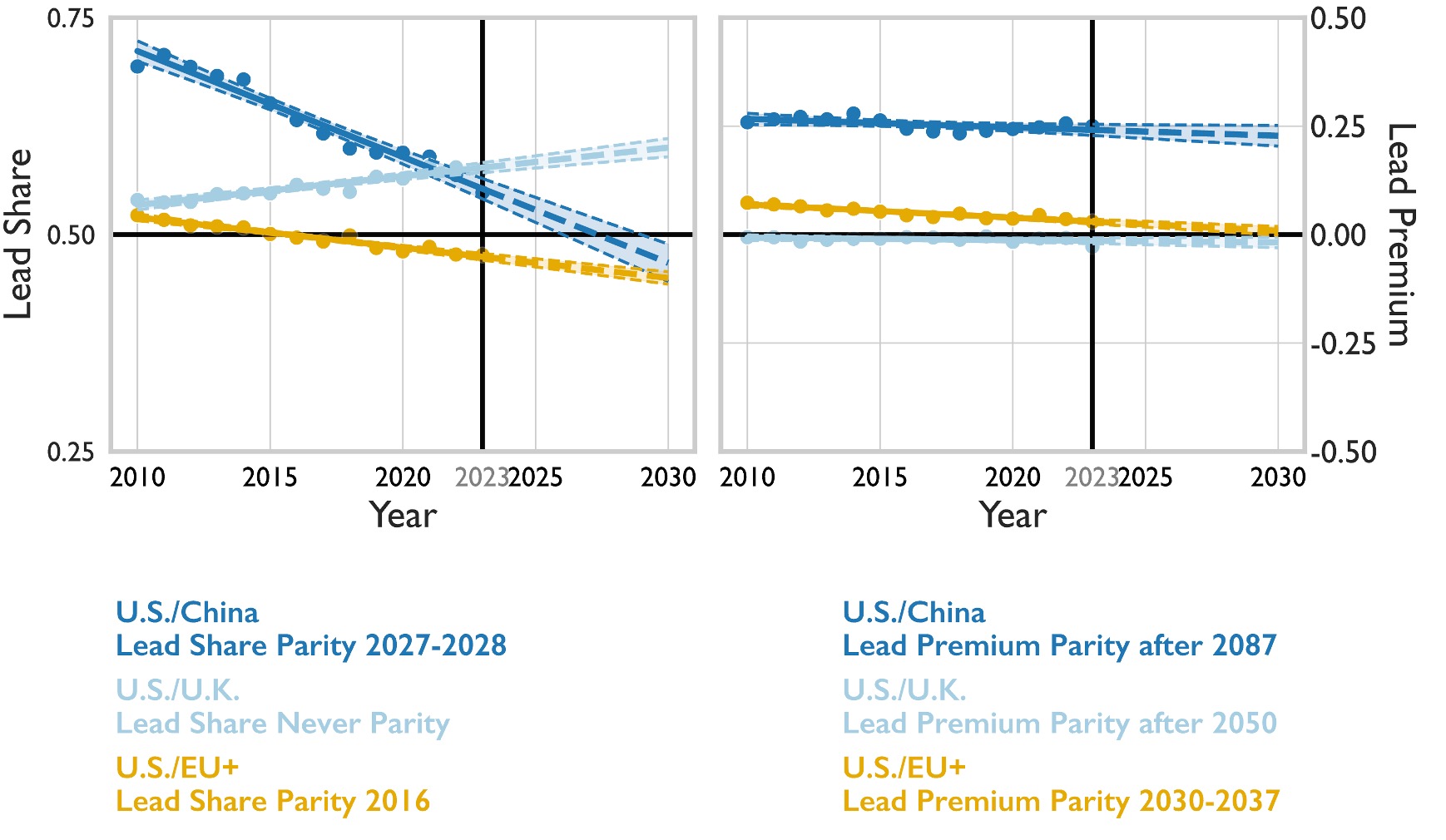}
\caption{US Leadership by Year}
\label{S5}
\end{figure}

\begin{figure}
\centering
\includegraphics[width=\textwidth]{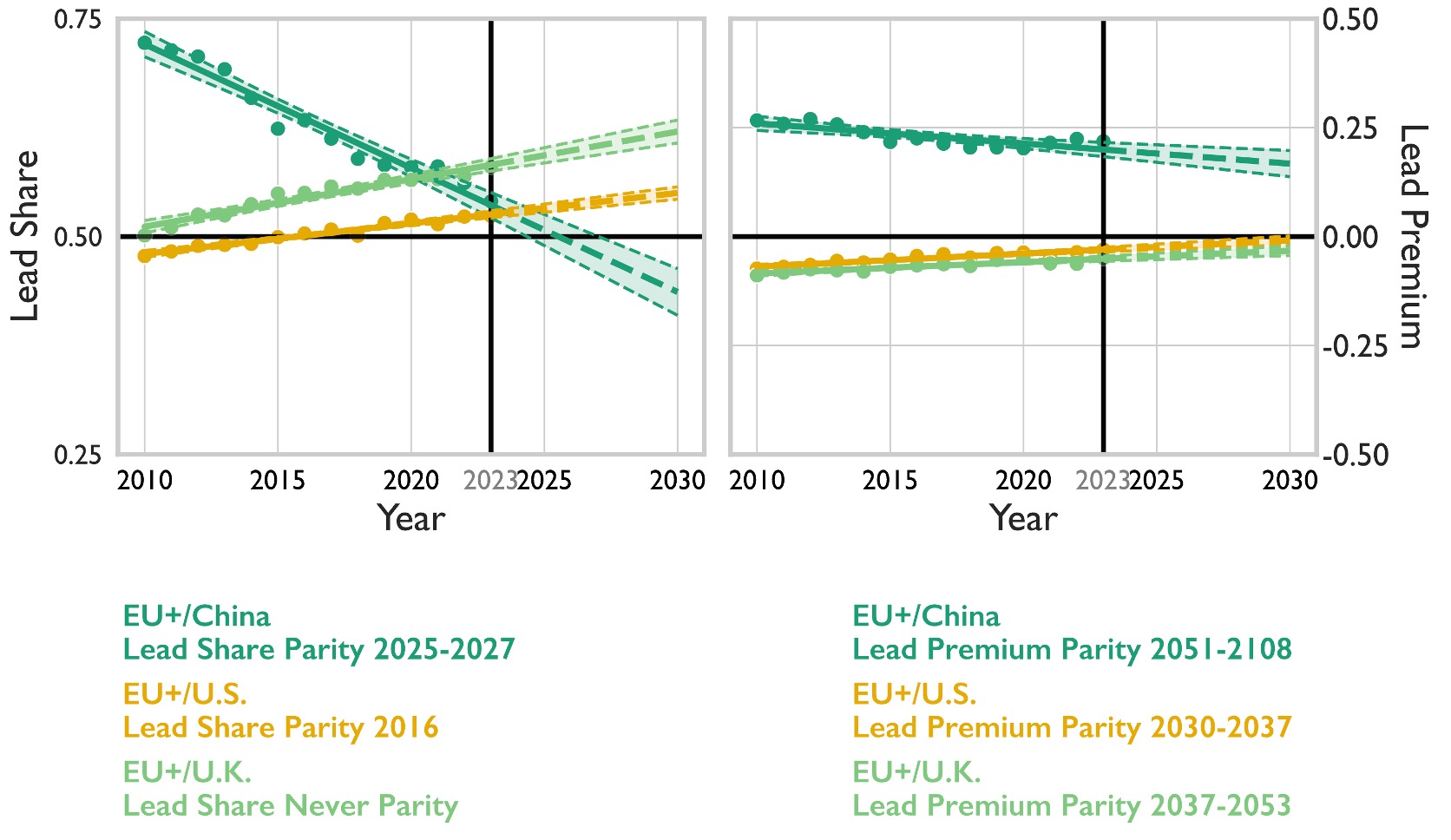}
\caption{EU Leadership by Year}
\label{S6}
\end{figure}

\begin{figure}
\centering
\includegraphics[width=\textwidth]{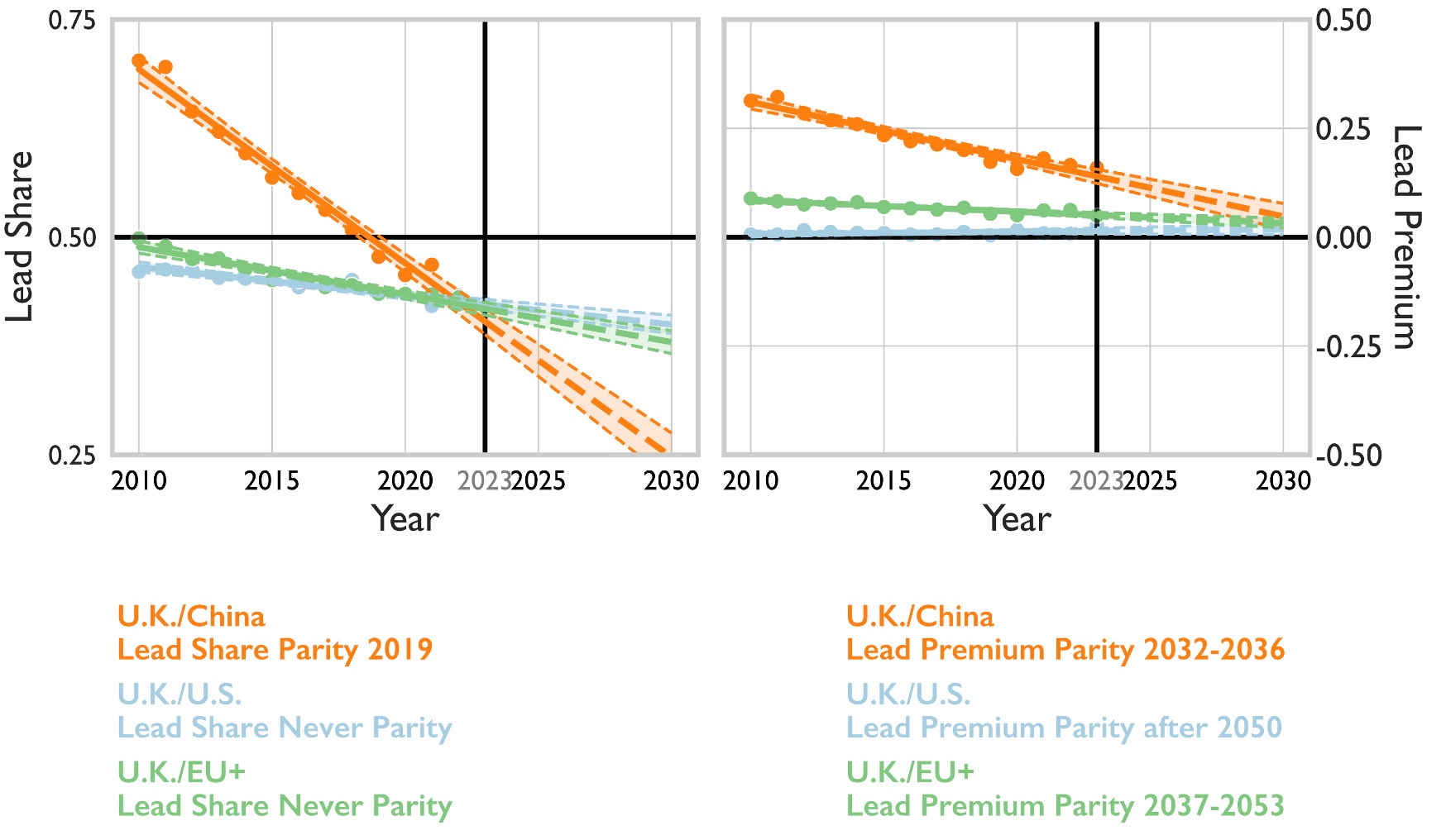}
\caption{UK Leadership by Year}
\label{S7}
\end{figure}

\begin{figure}
\centering
\scriptsize{\textbf{Advanced Communications:} advanced communications technology, communications technology, impacts to communications from long term changes in atmospheric condition, wireless communication, optical communication, 5g, internet of thing, satellite communication, network protocol, digital signal processing, telecommunication system, mobile network, broadband technologie, data transmission, communication security, machine to machine communication, radio frequency technologie,  cognitive radio network, voip, multimedia communication, network architecture, cyber physical system, immersive technology, virtual reality, augmented reality, mixed reality, extended reality, 360 degree video, haptic feedback, immersive experience, spatial computing, virtual environment, wearable technology, head mounted display, 3d modeling, virtual world, virtual presence. \textbf{Artificial Intelligence:} zero shot learning, artificial general intelligence, large language model, transfer learning, foundation model, deep reinforcement learning, human in the loop deep reinforcement learning, generative adversarial network, neural network, generative artificial intelligence, convolutional neural network, recurrent neural network, tensor processing unit, artificial neural network, ethics and responsible artificial intelligence, few shots learning, explainable artificial intelligence, one shot learning, deep learning, backpropagation, model evaluation, machine learning, adversarial machine learning, decision tree, random forest, support vector machine, regression, ensemble method, human machine teaming, robotic, unmanned system, autonomous system, sensor fusion, path planning, control system, human robot interaction, navigation system, supervised learning, unsupervised learning, semi supervised learning, multi agent system, natural language processing, data science, data analytic, data mining, classification, gradient descent, clustering, feature extraction, dimensionality reduction, hyperparameter tuning, computer vision. \textbf{Biotech:} biotechnology, bioeconomy, medical technology, biomanufacturing, genomic, bioremediation, synthetic biology, genetic sequencing, biopharmaceutical, biomechanic, tissue engineering, biomaterial, regenerative medicine, biomedical imaging, medical device, drug delivery, genetic engineering, bioinformatic, molecular biology, cell therapy, nanotechnology, bioprocessing, bioreactor, prosthetic, orthopedic engineering, bioengineering, biocompatibility, biosensor, neural engineering, immunotherapy, stem cell, biomedical signal processing, bioprinting, biomem, biomedical microelectromechanical system, precision medicine, cardiovascular engineering, clinical engineering, biophotonic, bioelectrical engineering, medical robotic, biomedical data analysis, pharmacokinetic, biomedical optic, biomolecular engineering, biochemical engineering, biomedical diagnostic, medical imaging technique, biostatistic, microfluidic, biofabrication, personalized medicine, biomedical instrumentation, functional genomic, computational biology, tissue mechanic, synthetic biology, rehabilitation engineering, biomedical ethic, biosignal processing. \textbf{Data Management and Security:} data management, data governance, data quality, database system, data security, data warehousing, data integration, metadata management, information lifecycle management, data privacy, data architecture, master data management, data lake, data virtualization, nosql database, sql, data cleaning, data migration, etl, extract, transform, load, distributed database, real time data processing, data lifecycle management, cloud hosted database, database management system, data integrity, secure data storage, data backup and recovery, cybersecurity, encryption, network security, information security, data privacy, cryptography, intrusion detection, firewall, malware analysis, risk management, access control, secure data transmission, phishing attack, data breach, blockchain, security protocol, threat intelligence, penetration testing, vulnerability assessment, cloud security, biometric security, identity management, secure authentication, compliance and regulation, advanced persistent threat, anomaly detection, security architecture, digital forensic, two factor authentication, zero trust security. \textbf{Disaster Resilience:} anthropogenic disaster prevention, climate resilience and adaptation, anthropogenic disaster mitigation, climate resiliency, natural disaster prevention, climate adaptation, natural disaster mitigation, disaster prevention, disaster mitigation, climate change, environmental degradation, industrial accident, pollution, deforestation, oil spill, nuclear accident, hazardous waste, urbanization, land degradation, air pollution, water contamination, soil erosion, overfishing, biodiversity loss, chemical spill, mining disaster, greenhouse gas emission, toxic waste, ozone depletion, drought, desertification, environmental health, resource depletion, ecosystem destruction, global warming, acid rain, pesticide contamination, overpopulation, habitat loss, wildfire, carbon footprint, e waste, landfill, ocean acidification, agricultural runoff, noise pollution, light pollution, plastic pollution, industrial emission, coral reef degradation, electromagnetic pollution, urban sprawl, invasive specie, non renewable energy, heavy metal contamination, environmental policy, ecological collapse, anthropogenic climate change. \textbf{Energy:} energy efficiency technology, battery management control, industrial efficiency technology, battery system integration, batterie, nuclear technology, battery management system, renewable energy, solar power, wind energy, energy storage, smart grid, energy efficiency, bioenergy, photovoltaic, battery technology, nuclear energy, hydroelectric power, thermal energy, electric vehicle, energy policy, fossil fuel, greenhouse gas emission, sustainable energy, energy conversion, geothermal energy, energy management, power generation, energy market, carbon capture, energy harvesting, offshore wind, grid integration, energy demand, hydrogen fuel cell, power system, climate change, energy security, energy modeling, oil and ga, distributed energy resource, energy economic, biomass, energy innovation, solar cell, wave energy, energy transition, environmental impact, clean energy, demand response, energy conservation, smart meter, tidal energy, energy regulation, energy auditing, power electronic, energy sector. \textbf{High Performance Computing:} high performance computing, parallel processing, high throughput computing, petascalegrid computing, cluster computing, floating point operations per second, supercomputer, parallel computing, parallel computing, distributed computing, supercomputing, cluster computing, multi core processing, high throughput computing, load balancing, gpu computing, simd, mimd, mpi, message passing interface, openmp, cuda, memory management, data intensive computing, cloud computing, heterogeneous computing, scheduling algorithm, scientific computing, grid computing, data parallelism, task parallelism, high performance networking, real time computing, exascale computing, storage system, file system, i/o performance, workflow management, parallel file system, advanced vector extension, avx, computational biology. \textbf{Materials Science:} materials science, composite 2d material, next generation material, sustainable material, carbon footprint, biodegradable material, nanomaterial, biomaterial, energy harvesting, organic electronic, lightweight material, photovoltaic, green building material, bio based polymer, polymer, composite, biomaterial, metallurgy, ceramic, superconductor, magnetic material, photovoltaic, nanotechnology, thin film, crystallography, carbon nanotube, graphene, alloy development, material characterization, corrosion, surface engineering, piezoelectric material, thermoelectric material, conductive material, mechanical propertie, polymer chemistry, nanoengineering, material synthesi, quantum dot, ferromagnetic material, materials testing, composite material, smart material, molecular dynamic, photonic crystal, electrochemistry, nanofabrication, coating, structural material, polymer science, materials processing, tribology, electronic material, optical material, amorphous material, nanocomposite, biodegradable material, nanowire, materials modeling, thermal propertie, self assembly, membrane science. \textbf{Quantum Tech:} quantum information science, quantum computer, quantum sensor, quantum technologie, quantum mechanic, quantum entanglement, quantum algorithm, quantum bit, qubit, superposition, quantum computing, quantum cryptography, quantum communication, quantum error correction, quantum key distribution, quantum teleportation, quantum gate, quantum circuit, quantum complexity theory, quantum simulation, quantum annealing, quantum decoherence, bell state, quantum fourier transform, no cloning theorem, quantum machine learning, quantum sensing, quantum metrology, quantum network, quantum state preparation, quantum information theory, quantum optic, quantum coherence, quantum control, quantum logic gate, quantum measurement, quantum noise, quantum programming language, quantum software, quantum hardware, quantum process tomography, quantum phase estimation, quantum chemistry, quantum topological state, quantum artificial intelligence, quantum optimization, quantum probability, quantum tunneling, quantum field theory, bloch sphere, quantum thermodynamic, quantum nonlocality, quantum entropy, quantum dynamic, heisenberg uncertainty principle. \textbf{Robotics and Advanced Manufacturing:} manufacturing: manufacturing processes, lean manufacturing, additive manufacturing, industrial engineering, supply chain management, quality control, production planning, mass production, computer aided manufacturing, cam, computer numerical control, cnc, robotic, 3d printing, process optimization, materials science, industrial robotic, smart manufacturing, six sigma, flexible manufacturing system, machining, assembly line, industrial iot, kaizen, just in time production, mrp, materials requirement planning, cad/cam integration, manufacturing execution system, precision engineering, sustainable manufacturing, injection molding, process control, computer aided design, eco friendly manufacturing, semiconductor, semiconductor material, silicon, gallium arsenide, transistor, integrated circuit, doping, semiconductor fabrication, microelectronic, photovoltaic, diode, moore's law, electron mobility, hole mobility, cmos technology, bipolar junction transistor, field effect transistor, semiconductor laser, optoelectronic, wafer fabrication, thin film transistor, semiconductor device, carrier concentration, p n junction. \textbf{Semiconductors:} semiconductor heterostructure, silicon carbide, organic semiconductor, photodetector, mem, microelectromechanical system, vlsi, very large scale integration, semiconductor packaging, electronic band structure, compound semiconductor, semiconductor metrology, charge carrier transport, sputtering, thermal oxidation, electron beam lithography, spintronic, superconducting semiconductor, semiconductor reliability, photolithography}
\caption{List of NSF Keywords by Critical Technology Area}
\label{S8}
\end{figure}

\begin{figure}
\centering
\scriptsize{\textbf{Advanced Communication:} Computer vision, Natural language processing, Speech recognition, Computer network, Knowledge management, Distributed computing, Computer graphics (images), Real-time computing, Multimedia, Computer architecture, World Wide Web, Human–computer interaction, Telecommunications, Communication. \textbf{Artificial Intelligence:} Artificial intelligence, Machine learning, Computer vision, Data science, Natural language processing, Speech recognition. \textbf{Biotech:} Genetics, Computational biology, Evolutionary biology, Biotechnology, Bioinformatics, Cancer research, Anatomy, Pharmacology, Physiology, Neuroscience, Medicinal chemistry, Internal medicine, Pathology, Surgery, Environmental health, Radiology, Cardiology, Nursing, Anesthesia, Gastroenterology, Physical therapy, Intensive care medicine, Pediatrics, Oncology, Family medicine, Gerontology, Dentistry, Audiology, Physical medicine and rehabilitation, Nuclear medicine, Dermatology, Gynecology, Medical education, Obstetrics, Medical emergency, Ophthalmology, Emergency medicine, Urology, Andrology, Traditional medicine, Orthodontics, Veterinary medicine, General surgery, Optometry, Biomedical engineering, Medical physics, Psychiatry, Clinical psychology, Molecular biology, Immunology, Virology. Data Management and Security: Operating system, Data science, Library science, Computer security, Database, Internet privacy. \textbf{Disaster Resilience:} Ecology, Molecular biology, Microbiology, Agronomy, Fishery, Animal science, Horticulture, Toxicology, Biological system, Food science, Agroforestry, Immunology, Virology, Organic chemistry, Environmental chemistry, Nuclear chemistry, Radiochemistry, Environmental economics, Agricultural economics, Environmental resource management, Environmental engineering, Geotechnical engineering, Water resource management, Soil science, Cartography, Physical geography, Forestry, Environmental planning, Environmental protection, Remote sensing, Seismology, Atmospheric sciences, Environmental regulation, Environmental health. \textbf{Energy:} Biochemistry, Natural resource economics, Nuclear engineering, Petroleum engineering, Nuclear physics, Photovoltaic system, Energy consumption, Wind power, Efficient energy use, Solar energy, Smart grid, Energy conservation, Hydropower, Hydroelectricity, Tidal power, Zero emission, Fuel cells, Cadmium telluride photovoltaics, Energy sector, Clean energy, Energy resources, Solar cell, Photovoltaics, Photovoltaic effect, Building-integrated photovoltaics, Renewable resource, Energy security, Sustainable energy, Energy planning, Non-renewable resource, Prosumer, Energy independence, Alternative energy, Battery (electricity) , Energy storage, Organic chemistry, Nuclear chemistry, Environmental economics. \textbf{High-Performance Computing:} Programming language, Library science, Computer security, Computer network, Information retrieval, Data mining, Database, Knowledge management, Distributed computing, Computer graphics (images), Parallel computing, Real-time computing, Multimedia, Internet privacy, Computer hardware, Computer architecture, Algorithm, Computational science, Embedded system, Software engineering, Simulation, Computer engineering. Materials Science: Chemistry, Chemical engineering, Biochemical engineering, Composite material, Nanotechnology, Metallurgy, Atomic physics, Biomedical engineering, Biochemistry. \textbf{Quantum Technology:} Nuclear magnetic resonance, Quantum electrodynamics, Hartree–Fock method, Variational method, Schrödinger's cat, Schrödinger equation, Isospin, Quantum tunneling, Quantum dot, Massless particle, Ground state, Density functional theory, Anomaly (physics), Dephasing, Pauli exclusion principle, Topological defect, Excited state, Hadron, Elementary particle, Quantum optics, Fermion, Perturbation theory (quantum mechanics), Asymmetry, Unitarity, Degenerate energy levels, Partition function (quantum field theory) , Topological quantum number, Fock space, Zero-point energy, Harmonic potential, Wave–particle duality, Dirac equation. Robotics and Advanced Manufacturing: Manufacturing engineering, Industrial engineering, Robotic arm, Servo, Mechanical system, Backlash, Servomotor, Automated guided vehicle, Harmonic drive, Driving simulator, Human arm, Haptic technology, Engineering design process, Production line, Product life-cycle management, Packaging and labeling, Conveyor system, Unit load, Packaging engineering, Work order, Kansei engineering, Group technology, Rapid prototyping, Stereolithography, Blank 3D printing, Bottleneck, Turnaround time, Capacity planning, Conformity assessment, Tying, Risk assessment, Contingency plan, Incident management, Life cycle costing, User experience design, Conceptual design, Interaction design, Design thinking, User-centered design, Design theory, Human engineering, Off the shelf, Computer Aided Design, Virtual prototyping, Production engineering, Assembly line, 3d printer, Computer-integrated manufacturing, Digital manufacturing, Nanomaterials, Batch production, Integrated logistics support, Critical quality attributes, Process validation. \textbf{Semiconductors:} Die (integrated circuit) , Application-specific integrated circuit, Gate count, Built-in self-test, Noise shaping, Dynamic range, Wide dynamic range, Digital signal processing, Static random-access memory, 8-bit, Very-large-scale integration, Circuit design, Integrated circuit design, Chemical deposition, Cascade, Silicon, Chemical vapor deposition, Electrode, Integrated circuit, Wafer, Doping, Band gap, Heterojunction, Charge carrier, Electron mobility, Superlattice, Wide-bandgap semiconductor, Rectangular potential barrier, Field effect, Semiconductor, Wavelength, Plasmon, Visible spectrum, Refractive index, Photonics, Ultraviolet, Photocurrent, Metamaterial, Waveguide, Photonic crystal, Terahertz radiation, Quantum efficiency, Photodetector, Transmittance, Photoconductivity, Photodiode, Photoelectric effect, Acceptor, Solid-state physics, Technology CAD, Telecommunications, Computer hardware, Computer engineering, Quantum electrodynamics
}
\caption{List of OpenAlex Concepts by Critical Technology Area}
\label{S9}
\end{figure}

\begin{figure}
\centering
\includegraphics[width=\textwidth]{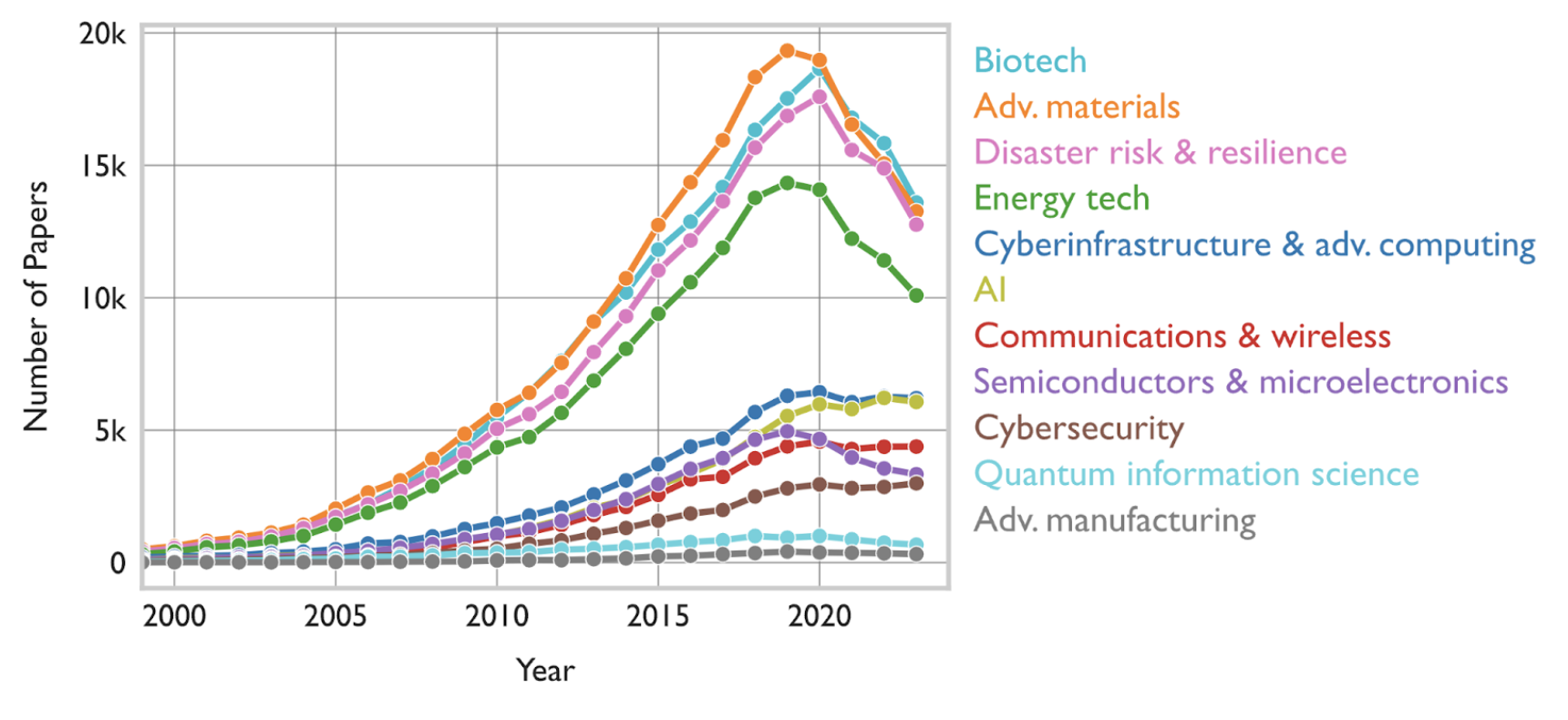}
\caption{Number of Collaborative Papers between Chinese and U.S. Scientists by Critical Technology Area}
\label{S10}
\end{figure}

\begin{figure}
\centering
\includegraphics[width=\textwidth]{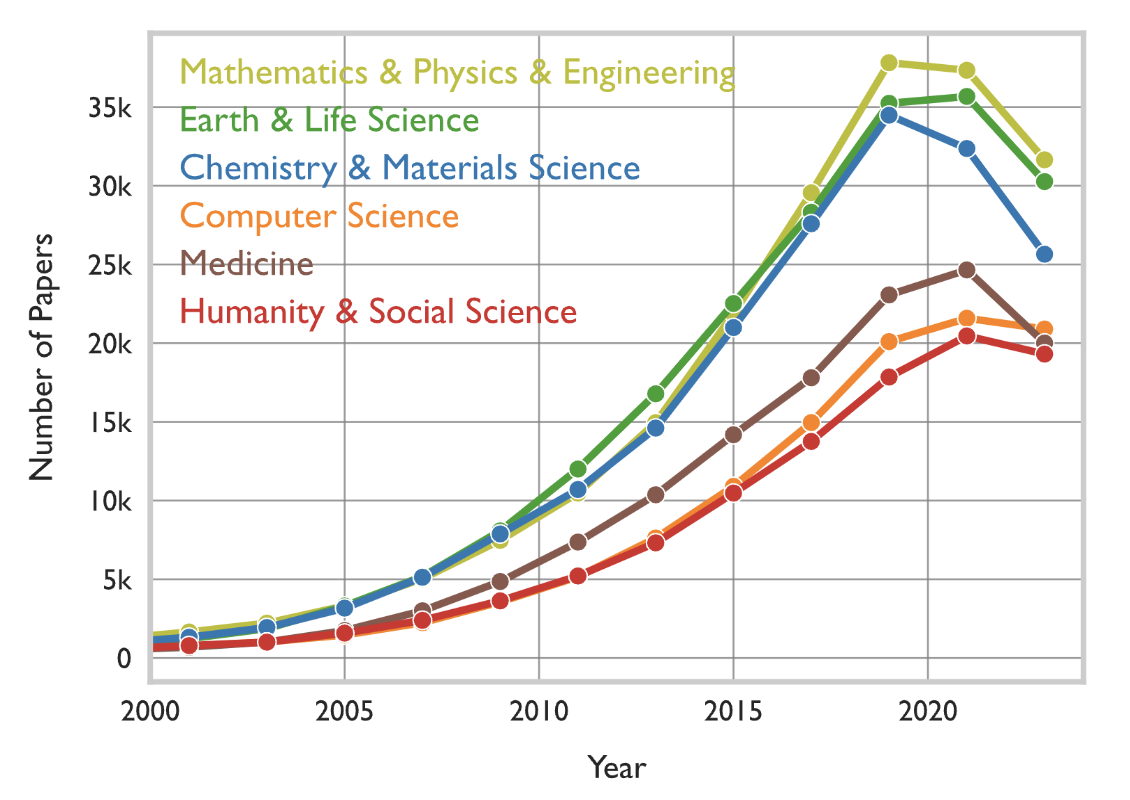}
\caption{Number of Collaborative Papers between Chinese and U.S. Scientists by Field}
\label{S11}
\end{figure}

\begin{figure}
\centering
\includegraphics[width=\textwidth]{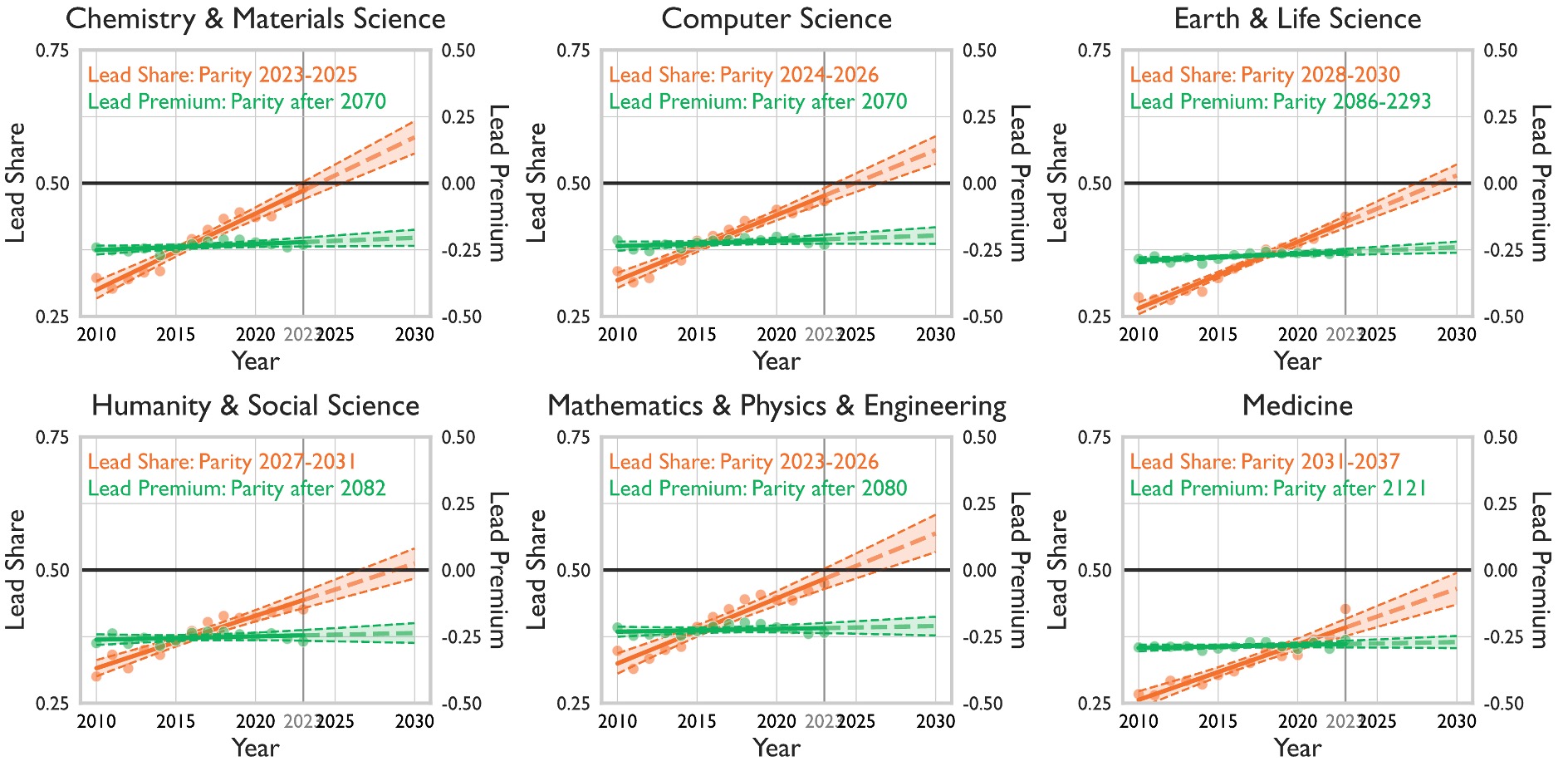}
\caption{Chinese Leadership in Collaborations with US Scientists by Technology Fields}
\label{S12}
\end{figure}

\begin{figure}
\centering
\includegraphics[width=\textwidth]{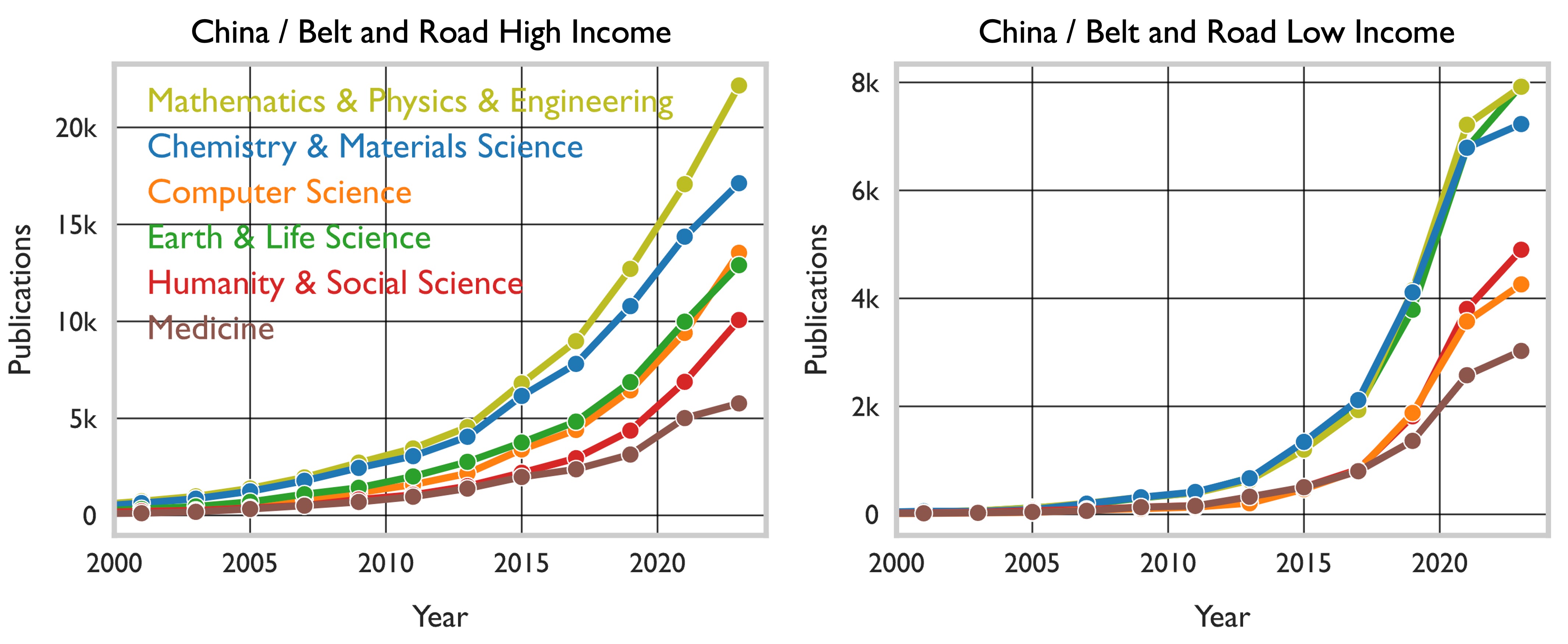}
\caption{Number of Publications by Scientists from China and Belt and Road Countries by Field}
\label{S13}
\end{figure}

\begin{figure}
\centering
\includegraphics[width=\textwidth]{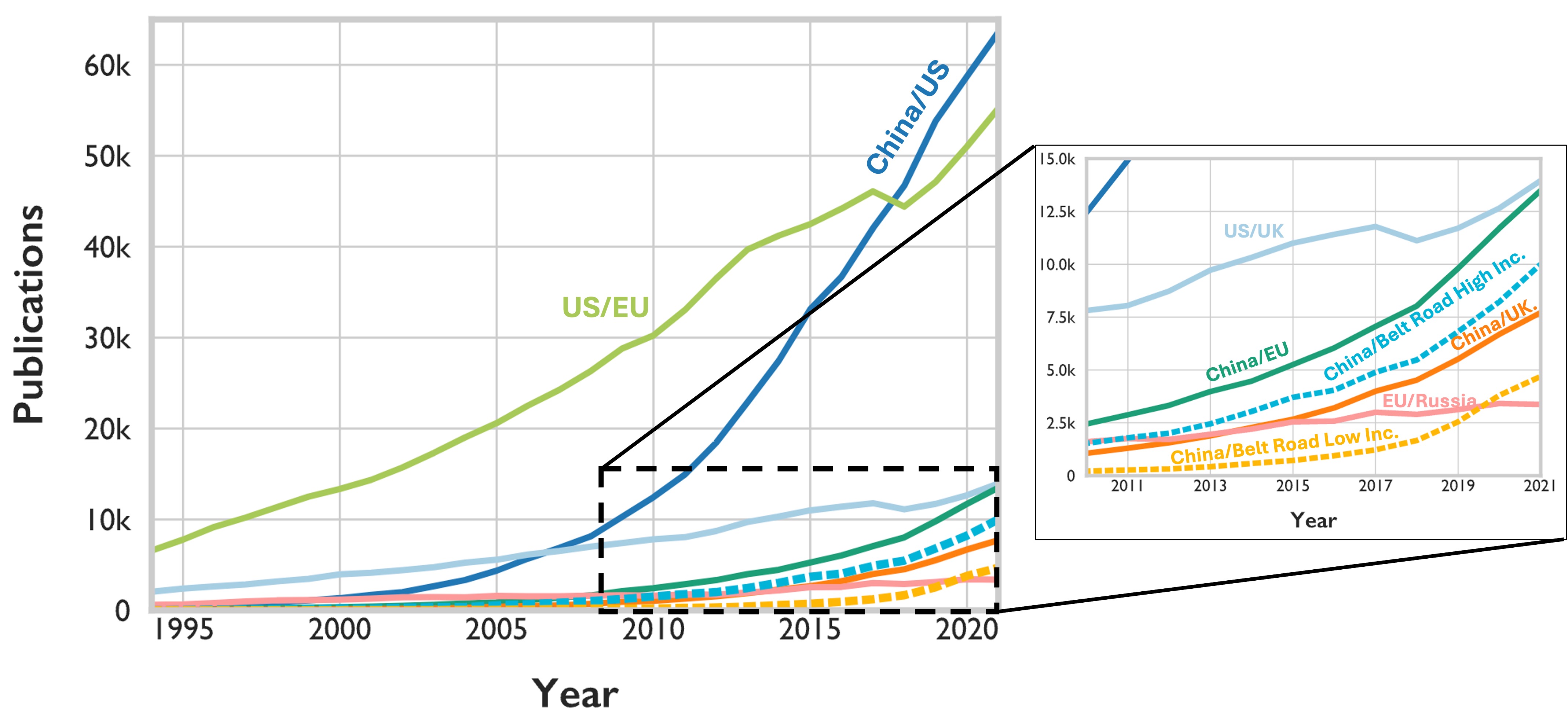}
\caption{Number of Papers by Region Pair Using MAG Data}
\label{S14}
\end{figure}

\begin{figure}
\centering
\includegraphics[width=\textwidth]{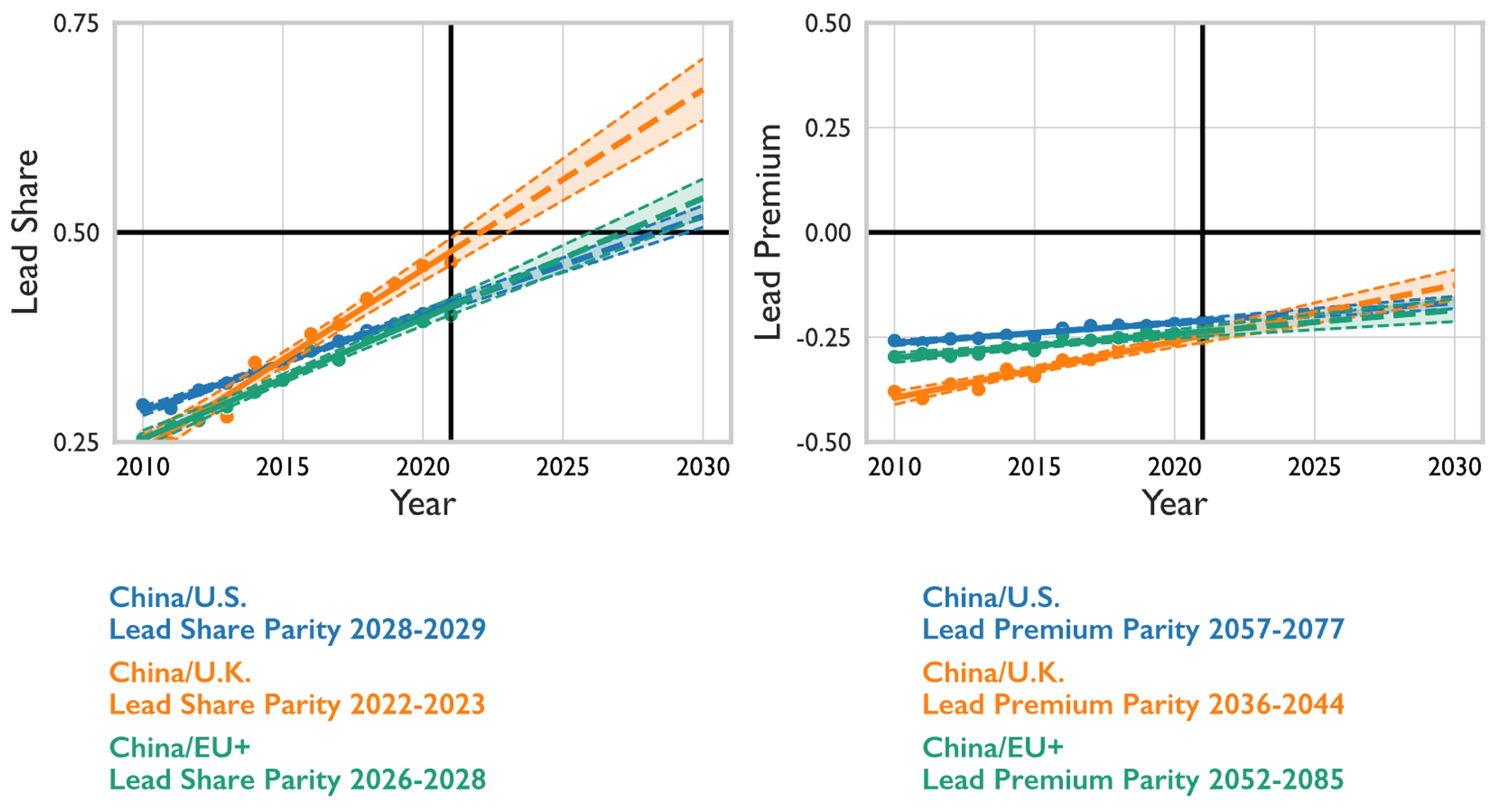}
\caption{Chinese Leadership with US, UK, and EU, MAG Data}
\label{S15}
\end{figure}

\begin{figure}
\centering
\includegraphics[width=\textwidth]{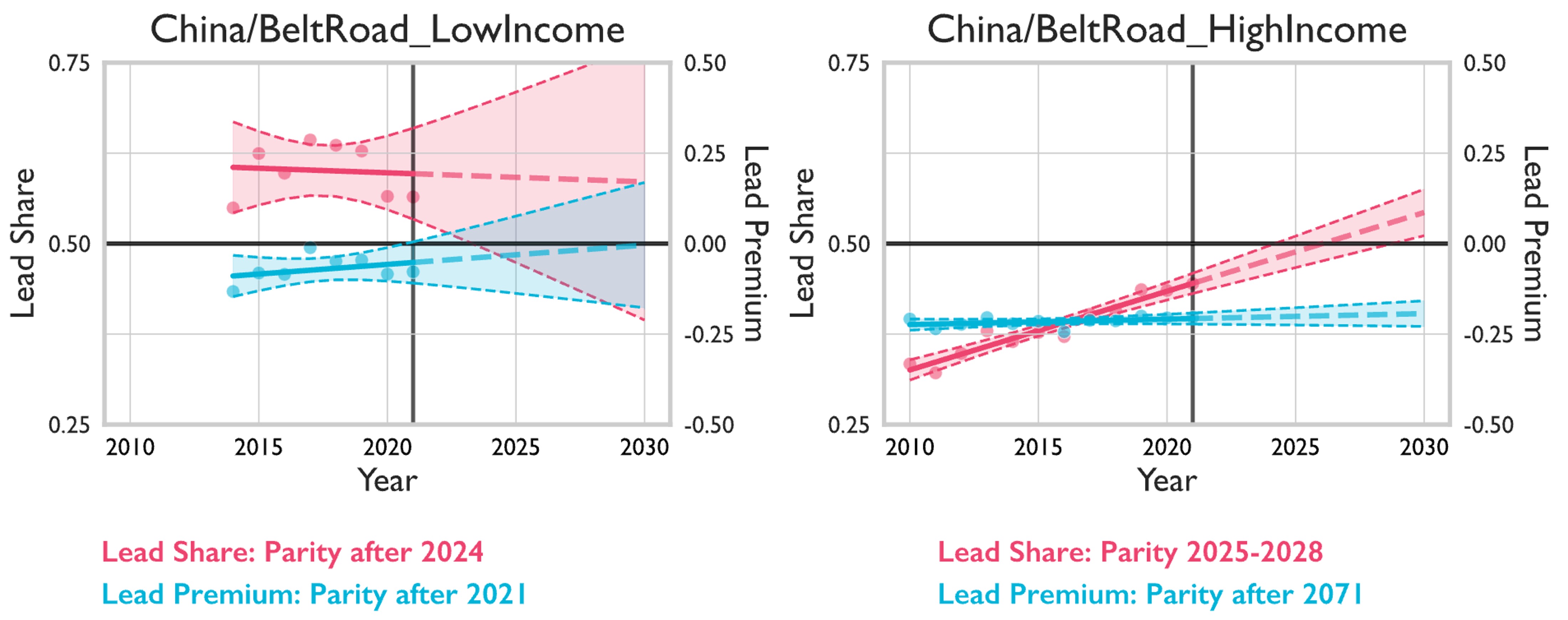}
\caption{Chinese Leadership with Belt and Road Countries, MAG Data}
\label{S16}
\end{figure}

\begin{table}[]
\small{
\begin{tabular}{
>{\columncolor[HTML]{EFEFEF}}l l
>{\columncolor[HTML]{EFEFEF}}l lll}
\cellcolor[HTML]{C0C0C0}\textbf{Country} & \cellcolor[HTML]{C0C0C0}\textbf{Region} & \cellcolor[HTML]{C0C0C0}\textbf{Country} & \cellcolor[HTML]{C0C0C0}\textbf{Region} & \cellcolor[HTML]{C0C0C0}\textbf{Country}      & \cellcolor[HTML]{C0C0C0}\textbf{Region} \\
Namibia                                  & Africa                                  & China                                    & China                                   & \cellcolor[HTML]{EFEFEF}United Arab Emirates  & Middle East                             \\
Angola                                   & Africa                                  & Japan                                    & East Asia                               & \cellcolor[HTML]{EFEFEF}Bahrain               & Middle East                             \\
Burundi                                  & Africa                                  & Korea, Rep.                              & East Asia                               & \cellcolor[HTML]{EFEFEF}Iraq                  & Middle East                             \\
Benin                                    & Africa                                  & South Korea                              & East Asia                               & \cellcolor[HTML]{EFEFEF}Iran, Islamic Rep.    & Middle East                             \\
Botswana                                 & Africa                                  & Solomon Islands                          & East Asia                               & \cellcolor[HTML]{EFEFEF}Kuwait                & Middle East                             \\
Democratic Republic of the Congo         & Africa                                  & Austria                                  & EU+                                     & \cellcolor[HTML]{EFEFEF}Lebanon               & Middle East                             \\
Congo, Rep.                              & Africa                                  & Belgium                                  & EU+                                     & \cellcolor[HTML]{EFEFEF}Oman                  & Middle East                             \\
Cote d'Ivoire                            & Africa                                  & Bulgaria                                 & EU+                                     & \cellcolor[HTML]{EFEFEF}Qatar                 & Middle East                             \\
Cameroon                                 & Africa                                  & Switzerland                              & EU+                                     & \cellcolor[HTML]{EFEFEF}Saudi Arabia          & Middle East                             \\
Cabo Verde                               & Africa                                  & Cyprus                                   & EU+                                     & \cellcolor[HTML]{EFEFEF}Syrian Arab Republic  & Middle East                             \\
Djibouti                                 & Africa                                  & Czech Republic                           & EU+                                     & \cellcolor[HTML]{EFEFEF}Yemen, Rep.           & Middle East                             \\
Algeria                                  & Africa                                  & Germany                                  & EU+                                     & \cellcolor[HTML]{EFEFEF}Albania               & Non-EU Eastern Europe                   \\
Egypt, Arab Rep.                         & Africa                                  & Denmark                                  & EU+                                     & \cellcolor[HTML]{EFEFEF}Belarus               & Non-EU Eastern Europe                   \\
Eritrea                                  & Africa                                  & Estonia                                  & EU+                                     & \cellcolor[HTML]{EFEFEF}Moldova               & Non-EU Eastern Europe                   \\
Ethiopia                                 & Africa                                  & Spain                                    & EU+                                     & \cellcolor[HTML]{EFEFEF}Montenegro            & Non-EU Eastern Europe                   \\
Gabon                                    & Africa                                  & Finland                                  & EU+                                     & \cellcolor[HTML]{EFEFEF}North Macedonia       & Non-EU Eastern Europe                   \\
Ghana                                    & Africa                                  & France                                   & EU+                                     & \cellcolor[HTML]{EFEFEF}Serbia                & Non-EU Eastern Europe                   \\
Gambia, The                              & Africa                                  & Greece                                   & EU+                                     & \cellcolor[HTML]{EFEFEF}Turkey                & Non-EU Eastern Europe                   \\
Guinea                                   & Africa                                  & Croatia                                  & EU+                                     & \cellcolor[HTML]{EFEFEF}Ukraine               & Non-EU Eastern Europe                   \\
Equatorial Guinea                        & Africa                                  & Hungary                                  & EU+                                     & \cellcolor[HTML]{EFEFEF}Australia             & Oceania                                 \\
Guinea-Bissau                            & Africa                                  & Ireland                                  & EU+                                     & \cellcolor[HTML]{EFEFEF}Cook Islands          & Oceania                                 \\
Kenya                                    & Africa                                  & Italy                                    & EU+                                     & \cellcolor[HTML]{EFEFEF}Fiji                  & Oceania                                 \\
Comoros                                  & Africa                                  & Lithuania                                & EU+                                     & \cellcolor[HTML]{EFEFEF}Micronesia, Fed. Sts. & Oceania                                 \\
Liberia                                  & Africa                                  & Luxembourg                               & EU+                                     & \cellcolor[HTML]{EFEFEF}Kiribati              & Oceania                                 \\
Lesotho                                  & Africa                                  & Latvia                                   & EU+                                     & \cellcolor[HTML]{EFEFEF}Niue                  & Oceania                                 \\
Libya                                    & Africa                                  & Malta                                    & EU+                                     & \cellcolor[HTML]{EFEFEF}New Zealand           & Oceania                                 \\
Morocco                                  & Africa                                  & Netherlands                              & EU+                                     & \cellcolor[HTML]{EFEFEF}Papua New Guinea      & Oceania                                 \\
Madagascar                               & Africa                                  & Norway                                   & EU+                                     & \cellcolor[HTML]{EFEFEF}Tonga                 & Oceania                                 \\
Mali                                     & Africa                                  & Poland                                   & EU+                                     & \cellcolor[HTML]{EFEFEF}Vanuatu               & Oceania                                 \\
Mauritania                               & Africa                                  & Portugal                                 & EU+                                     & \cellcolor[HTML]{EFEFEF}Samoa                 & Oceania                                 \\
Mauritius                                & Africa                                  & Romania                                  & EU+                                     & \cellcolor[HTML]{EFEFEF}Russian Federation    & Russia                                  \\
Malawi                                   & Africa                                  & Sweden                                   & EU+                                     & \cellcolor[HTML]{EFEFEF}Bangladesh            & South Asia                              \\
Mozambique                               & Africa                                  & Slovenia                                 & EU+                                     & \cellcolor[HTML]{EFEFEF}Brunei Darussalam     & South Asia                              \\
Niger                                    & Africa                                  & Slovakia                                 & EU+                                     & \cellcolor[HTML]{EFEFEF}Indonesia             & South Asia                              \\
Nigeria                                  & Africa                                  & Slovak Republic                          & EU+                                     & \cellcolor[HTML]{EFEFEF}India                 & South Asia                              \\
Rwanda                                   & Africa                                  & Antigua and Barbuda                      & Latin America                           & \cellcolor[HTML]{EFEFEF}Cambodia              & South Asia                              \\
Seychelles                               & Africa                                  & Argentina                                & Latin America                           & \cellcolor[HTML]{EFEFEF}Lao PDR               & South Asia                              \\
Sudan                                    & Africa                                  & Barbados                                 & Latin America                           & \cellcolor[HTML]{EFEFEF}Sri Lanka             & South Asia                              \\
Sierra Leone                             & Africa                                  & Bolivia                                  & Latin America                           & \cellcolor[HTML]{EFEFEF}Myanmar               & South Asia                              \\
Senegal                                  & Africa                                  & Brazil                                   & Latin America                           & \cellcolor[HTML]{EFEFEF}Maldives              & South Asia                              \\
Somalia                                  & Africa                                  & Chile                                    & Latin America                           & \cellcolor[HTML]{EFEFEF}Malaysia              & South Asia                              \\
South Sudan                              & Africa                                  & Colombia                                 & Latin America                           & \cellcolor[HTML]{EFEFEF}Philippines           & South Asia                              \\
Chad                                     & Africa                                  & Costa Rica                               & Latin America                           & \cellcolor[HTML]{EFEFEF}Pakistan              & South Asia                              \\
Togo                                     & Africa                                  & Cuba                                     & Latin America                           & \cellcolor[HTML]{EFEFEF}Singapore             & South Asia                              \\
Tunisia                                  & Africa                                  & Dominica                                 & Latin America                           & \cellcolor[HTML]{EFEFEF}Thailand              & South Asia                              \\
Tanzania                                 & Africa                                  & Dominican Republic                       & Latin America                           & \cellcolor[HTML]{EFEFEF}Timor-Leste           & South Asia                              \\
Uganda                                   & Africa                                  & Ecuador                                  & Latin America                           & \cellcolor[HTML]{EFEFEF}Vietnam               & South Asia                              \\
South Africa                             & Africa                                  & Grenada                                  & Latin America                           & \cellcolor[HTML]{EFEFEF}United Kingdom        & U.K.                                    \\
Zambia                                   & Africa                                  & Guyana                                   & Latin America                           & \cellcolor[HTML]{EFEFEF}United States         & U.S.                                    \\
Zimbabwe                                 & Africa                                  & Jamaica                                  & \multicolumn{2}{l}{Latin America}                                                       &                                         \\
Afghanistan                              & Central Asia                            & Mexico                                   & \multicolumn{2}{l}{Latin America}                                                       &                                         \\
Armenia                                  & Central Asia                            & Nicaragua                                & \multicolumn{2}{l}{Latin America}                                                       &                                         \\
Azerbaijan                               & Central Asia                            & Panama                                   & \multicolumn{2}{l}{Latin America}                                                       &                                         \\
Georgia                                  & Central Asia                            & Peru                                     & \multicolumn{2}{l}{Latin America}                                                       &                                         \\
Kyrgyz Republic                          & Central Asia                            & Suriname                                 & \multicolumn{2}{l}{Latin America}                                                       &                                         \\
Kazakhstan                               & Central Asia                            & El Salvador                              & \multicolumn{2}{l}{Latin America}                                                       &                                         \\
Mongolia                                 & Central Asia                            & Trinidad and Tobago                      & \multicolumn{2}{l}{Latin America}                                                       &                                         \\
Nepal                                    & Central Asia                            & Uruguay                                  & \multicolumn{2}{l}{Latin America}                                                       &                                         \\
Tajikistan                               & Central Asia                            & Venezuela, RB                            & \multicolumn{2}{l}{Latin America}                                                       &                                         \\
Turkmenistan                             & Central Asia                      \\
Uzbekistan                               & Central Asia                      \\
\end{tabular}
}
\caption{Countries by Region of Geopolitical Significance}
\label{T1}
\end{table}

\FloatBarrier
